\documentclass[useAMS,usenatbib]{mn2e}
\usepackage{epsfig}
\usepackage{aas_macros}     % for journal abbreviations in references
\usepackage{amsfonts}

\usepackage[usenames]{color}
\usepackage{url}

\newcommand{\mpl}{M_{\rm Pl}}
\newcommand{\rd}{{\rm d}}

\newcommand{\tc}{\textcolor{blue}}

\newcommand{\remove}[1]{}

\def\ie{{\frenchspacing\it i.e.}}
\def\eg{{\frenchspacing\it e.g.}}

\def\be{\begin{equation}}
\def\ee{\end{equation}}
\def\ba{\begin{eqnarray}}
\def\ea{\end{eqnarray}}

\title[Redshift space distortions in $f(R)$ gravity ]
{Redshift space distortions in $f(R)$ gravity}
\author[Elise~Jennings, Carlton~M.~Baugh, Baojiu~Li,  Gong-Bo~Zhao, Kazuya Koyama]
{Elise~Jennings$^{1,2}$\thanks{E-mail: ejennings@kicp.uchicago.edu}, Carlton~M.~Baugh$^{3}$, Baojiu~Li$^{3}$,
Gong-Bo~Zhao$^{4,5}$, 
Kazuya Koyama$^{4}$\\
$^{1}$ The Kavli Institute for Cosmological Physics, University of Chicago, 5640 South Ellis Avenue, Chicago, IL 60637, U. S.\\
$^{2}$ The Enrico Fermi Institute, University of Chicago, 5640 South Ellis Avenue, Chicago, IL 60637, U. S.\\
$^{3}$ Institute of Computational Cosmology, Department of Physics, Durham University, South Road, Durham DH1 3LE, U. K.\\
$^{4}$ Institute of Cosmology \& Gravitation, University of Portsmouth, Portsmouth, PO1 3FX, U. K.\\
$^{5}$ National Astronomy Observatories, Chinese Academy of Science, Beijing, 100012, P.R.China}

\begin{document}

\date{\today}

\pagerange{\pageref{firstpage}--\pageref{lastpage}} \pubyear{2012}

\maketitle

\label{firstpage}

\begin{abstract}
We use large volume N-body simulations
to predict the clustering of dark matter in redshift space
in $f(R)$ modified gravity cosmologies.
This is the first time that the nonlinear matter and
velocity fields have been resolved to such a high level
of accuracy over a broad range of scales in this class of models.
We find  significant deviations from the clustering signal
in standard gravity, with an enhanced boost in power on large scales
and stronger damping on small scales in the $f(R)$ models
compared to GR at redshifts $z<1$.
We measure the velocity divergence ($P_{\theta \theta}$) and
matter ($P_{\delta \delta}$) power spectra and
find a large deviation in the ratios
$\sqrt{P_{\theta \theta}/P_{\delta \delta}}$ and $P_{\delta
\theta}/P_{\delta \delta}$ between the $f(R)$ models and GR for 
$0.03<k/(h/$Mpc$)<0.5$. In linear theory these ratios equal
the growth rate of structure on large scales.
Our results show that the simulated ratios agree with the growth rate
 for each cosmology (which is scale dependent in the case of modified
gravity) only for extremely large scales, $k<0.06h/$Mpc at $z=0$.
The velocity power spectrum is substantially different
in the $f(R)$ models compared to GR, suggesting that this observable
is a sensitive probe of modified gravity.
We demonstrate how to extract the matter and velocity power
spectra from the 2D redshift space power spectrum, $P(k,\mu)$,
and can recover the nonlinear matter power spectrum to within
a few percent for $k<0.1h/$Mpc.
However, the model fails to
describe the shape of the 2D power spectrum demonstrating that an improved model 
 is necessary in order to reconstruct the velocity power spectrum accurately.
 The same model can match the
monopole moment to within 3\% for GR and 10\% for the $f(R)$
cosmology at $k<0.2 h/$Mpc at $z=1$. 
Our results suggest that the extraction
of the velocity power spectrum from future galaxy surveys is a promising method to
constrain deviations from GR.
\end{abstract}


\section{Introduction}

\label{sect:intro}

The  clustering of galaxies on different scales
is a key observational tool in the quest to explain the current accelerating expansion of the Universe
\citep{Percival:2007yw, Guzzo, betal2010,  betal2011,BOSS,DES,Schlegel2009,Laureijs, LSST,
Green}. The accelerating expansion may be the result of a dark energy component which behaves as a repulsive form of gravity or it may be that Einstein's 
theory of gravity breaks down on cosmological scales \citep[see e.g. ][]{bz2008,Weinberg:2012es}.
For a given cosmology with a smooth dark energy component, a measurement of the expansion history gives a prediction for the growth rate of structure. 
Independent measurements of the growth rate can be obtained by measuring the clustering of galaxies in redshift space, where peculiar 
velocities distort the clustering signal along the line of sight. 
By testing the consistency between the measured growth rate and the prediction from the expansion history, it is possible to 
constrain models of modified gravity and to distinguish them from a smooth dark energy component \citep[see e.g.][]{mhh2009, v2012}.
In this paper we measure the anisotropic power spectrum in redshift space from 
large volume N-body simulations of $f(R)$ modified gravity and general relativity (GR) cosmologies.

%Alternative models which explain the accelerating expansion of the Universe,
% without the use of a dark energy component, propose a modification of gravity on cosmological scales.
The $f(R)$ class of models can mimic the effect of a cosmological constant and is set up by modifying the Einstein-Hilbert action with
 an arbitrary function of the Ricci scalar, $R$ \citep[see e.g.][]{carroll2003, no2003}. A key 
feature of these models is the existence of a \lq fifth force\rq,  due to  an extra
propagating scalar field. 
Departures from general relativity on small scales are highly constrained by solar system tests \citep[e.g.][]{will2006}. 
As a result viable $f(R)$ theories must exhibit a screening mechanism, the so-called chameleon effect \citep{kw2004}, 
whereby standard gravity is recovered in 
high density environments. 
The range of the fifth force depends nonlinearly
on the local curvature and as a result will change with redshift.
The impact of the chameleon mechanism on the
matter and velocity fields can only be fully investigated using $N$-body simulations.
Any deviations from standard gravity 
will depend on the choice of the function $f(R)$ and the parameter values adopted.

In this paper we consider the $f(R)$ model  proposed by \citet{hs2007}. This modified gravity model has been
incorporated into  $N$-body simulations and studied by several authors \citep{oyaizu2008,olh2008,sloh2009,svh2009,lssh2010,fsh2010,zlk2011,lh2011,gil2011}
A variety of computational box sizes from 64 to 400 Mpc$/h$ on a side have been used.
In this work we make use of large volume, $L_{\tiny \mbox{box}}=1.5$Gpc$/h$ and  1Gpc$/h$, modified gravity simulations  
using the $N$-body code of \citet{lztk2012}.  
The large volume of these simulations allow us to study the impact of unique features of modified gravity, such as the
 scale dependent enhanced gravitational force, on the clustering signal in redshift space.

\begin{figure}
{\epsfxsize=8.truecm
\epsfbox[53 361 371 616]
{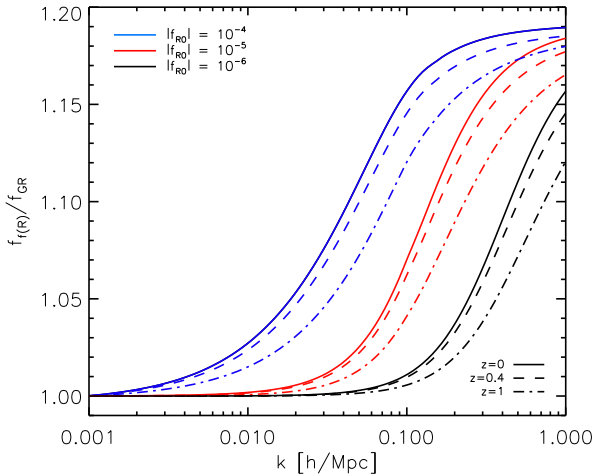}
}
\caption{
The ratio of the linear growth rate  in the F4 (blue), F5 (red) and
F6 (black) cosmologies to that in $\Lambda$CDM. The ratios are shown at $z=0$ (solid), $z=0.4$ (dashed) and $z=1$ (dot dashed).
(See Section 2.2 for the description of
the modified gravity models.)
\label{growthrate}}
\end{figure}

Galaxy redshift surveys allow us to study the 3D spatial distribution of galaxies and clusters.
In addition to the Hubble flow, galaxies have peculiar velocities, due to local inhomogenities in the density field, which distort the measured distances.
Measuring the anisotropic distortions in the galaxy clustering pattern in redshift space 
constrains $\beta = f/b$, where 
$b$ is the galaxy bias factor and $f$ is the logarithmic derivative of the linear growth rate of structure, which
is scale independent in the case of general relativity.
This effect was first described by  \citet{kaiser1987} using linear
perturbation theory where the matter power spectrum
in redshift space can be expressed as a function of the power spectrum in real space and $\beta$. 
Several authors  have extended this linear model to quasi linear scales by e.g. 
including nonlinear velocity terms \citep{scoccimarro2004, matsubara2008, pw2009,t2010} or by 
considering a phase space distribution function approach \citep{seljak2011} or into the 
  nonlinear regime by including the contribution of
peculiar velocities on small scales \citep[e.g.][]{p1976,pd1994,rw2011}.

Modelling the clustering of the dark matter and galaxies in redshift space is extremely challenging. 
Most  models contain free parameters such as the linear bias, which quantifies the 
difference in clustering between the dark matter and galaxies on large scales, 
and the velocity dispersion due to incoherent motions 
on small scales \citep[see e.g.][]{o2011}. These parameters must be included when
fitting any model and can weaken the constraints on the growth rate.

Many redshift space distortion models which are currently used are 
only accurate for a limited range of scales or for galaxies with a particular linear 
bias \citep[e.g.][]{scoccimarro2004,rw2011}.  
\citet{scoccimarro2004} proposed a simple quasi-linear model which includes 
the nonlinear velocity power spectrum. By comparing with measurements from $N$-body simulations, \citet{jbp2011a,jbp2011b} 
showed that this model 
performed better than commonly
used models (which we discuss in Section 3) and
is accurate on scales $k<0.3h/$Mpc and can recover the linear growth rate to within a few percent. 
The nonlinear velocity terms in this model may be obtained using either a fitting formula calibrated against 
$N$-body simulations or from perturbation theory \citep{matsubara2008,jbp2011a}.

The WiggleZ Dark Energy Survey \citep{betal2011} recently measured the growth rate at $z= 0.78$ to be $f= 0.70 \pm 0.08$ using redshift-space distortions in the galaxy power spectrum.
Measurements of the linear growth
rate are degenerate with the bias or clustering amplitude in the power spectra and so contraints on the 
growth rate are often quoted as contraints on $f\sigma_8$, where $\sigma_8$ is the {\it rms} variance in the linear matter
power spectra smoothed in spheres of radius 8 Mpc$/h$ \citep{pw2009}.
 The 6dF Galaxy Survey  modelled 
the 2D galaxy correlation function and obtained a  low redshift measurement of the growth rate,
$f\sigma_8= 0.423 \pm 0.055 $, at an effective redshift of $z_=0.067$ \citep{beutler2012}. 
Recent measurements from the SDSS III BOSS survey  found 
${\rm d}\sigma_8/{\rm d ln}a = 0.43 \pm 0.069 $ at an effective redshift of $z=0.57$ \citep{retal2012}. 
All of these results are consistent with the $\Lambda$CDM
model and standard gravity. Current surveys do not have sufficient 
precision to rule out viable modified gravity models such as the $f(R)$ models considered in this paper.
Future galaxy redshift surveys, such as the ESA's EUCLID mission \citep{Laureijs} and the ground-based stage IV dark energy experiment, BigBOSS \citep{Schlegel2009}, 
aim to measure the growth rate to within 2\%, which 
will place significant constraints on currently allowed modified gravity models.

A key feature of redshift space distortion          
 models is that the linear growth rate is assumed to be scale independent.
This assumption is not true for the modified gravity cosmology considered in this paper (see Fig. \ref{growthrate}).
In addition, several models of redshift space distortions suffer from systematic biases when fitting for a scale independent
growth rate over a range of scales \citep[see Figure 5 in ][]{jbp2011b}.
In order to avoid assuming a specific scale dependence for the growth rate,
 we instead 
focus on recovering the velocity and matter power spectra as a function of scale using the full
two dimensional redshift space power spectrum. 
This approach makes use of the full 2D power spectrum measured from a survey and the extracted
matter and velocity power spectra could  be compared to predictions from the standard cosmological model.

In this paper we measure the power spectrum in redshift space from a suite of large volume 
simulations of $f(R)$ cosmologies \citep{lztk2012}. This is the first time that predictions for the redshift space 
clustering  in $f(R)$ modified gravity models have been presented.
The  resolution of our simulations allows us to accurately resolve the nonlinear matter and velocity fields
and quantify the deviations from a model of general relativity. Here, we restrict our study 
to the clustering of the dark matter. 
We examine the difference between the velocity power spectra in each cosmology
and its importance in modelling the redshift space clustering signal in both the standard and modified gravity
model. We test how well quasi linear models for the redshift space distortions describe the amplitude and shape of
the measured power spectrum.
A follow up paper will examine the redshift space 
distortions in the clustering of halos as well as testing  nonlinear models for redshift space distortions \citep[see also 
e.g.][for a recent study of redshift space distortions in interacting dark energy models]{mbm2012}.

This paper is organised as follows: In Section~\ref{sect:fr} we discuss the $f(R)$ modified gravity cosmological model and  describe the
 $N$-body simulations used in this paper. In Section~\ref{RSD} we review the theory of redshift space distortions and present the models which will be tested.
The main results of the paper are presented in Section~\ref{RESULTS}.
Measurements of the  redshift space power spectra for both general relativity and the $f(R)$ models  are presented in Section~\ref{section_pk}.
In Section~\ref{Velocity} we present the velocity power spectrum measured from the simulations. 
Using a quasi linear model for the redshift space power spectrum we attempt to extract both the matter and velocity power spectra from the 
two dimensional redshift space power spectrum in Section~\ref{reconstructing}. In Section~\ref{moments} 
we examine how well the moments of the redshift space power spectrum can be recovered
using this quasi linear model. Our conclusions are presented in Section~\ref{CONCLUSIONS}.
Throughout the paper we shall use the unit $c=1$ and metric convention $(+,-,-,-)$. Greek letters $\mu,\nu,\cdots$ run over $0,1,2,3$ and Latin letters $i,j,k,\cdots$ run over $1,2,3$.

\section{$f(R)$ cosmologies}

\label{sect:fr}
This section gives the theoretical background for the 
modified gravity model considered in this paper.
We outline $f(R)$ cosmologies in Section 2.1, explain the chameleon mechanism in Section 2.2 and describe 
the $N$-body code and simulations in Section 2.3.

\subsection{The $f(R)$ gravity model}

\label{subsect:fr}

The $f(R)$ gravity model is a straightforward generalisation of GR: the Ricci scalar, $R$, in 
the Einstein-Hilbert action, $S$, is replaced with an algebraic function, $f(R)$ \citep[see \eg,][for recent reviews]{sf2010, dt2010}:
\begin{eqnarray}\label{eq:fr_action}
S &=& \int{\rm d}^4x\sqrt{-g}\left\{\frac{\mpl^2}{2}\left[R+f(R)\right]+\mathcal{L}_m\right\},
\end{eqnarray}
in which $\mpl$ is the Planck mass, $\mpl^{-2}=8\pi G$ with $G$ being Newton's constant, $g$ is the determinant of the metric $g_{\mu\nu}$ and $\mathcal{L}_m$ is the Lagrangian density for matter fields (photons, neutrinos, baryons and cold dark matter). By specifying the functional form of $f(R)$ one specifies the $f(R)$ gravity model.

Varying the action defined in Eq.~(\ref{eq:fr_action}) with respect to the metric $g_{\mu\nu}$ yields the modified Einstein equation
\begin{eqnarray}\label{eq:fr_einstein}
G_{\mu\nu} + f_RR_{\mu\nu} -\left(\frac{1}{2}f-\Box f_R\right)g_{\mu\nu}-\nabla_\mu\nabla_\nu f_R = 8\pi GT^m_{\mu\nu},
\end{eqnarray}
in which $G_{\mu\nu}\equiv R_{\mu\nu}-\frac{1}{2}g_{\mu\nu}R$ is the Einstein tensor, $f_R\equiv \rd f/\rd R$, $\nabla_{\mu}$ the covariant derivative compatible with 
the metric $g_{\mu\nu}$, $\Box\equiv\nabla^\alpha\nabla_\alpha$ and $T^m_{\mu\nu}$ is the energy momentum tensor for matter. One can consider Eq.~(\ref{eq:fr_einstein}) as a fourth-order differential equation, or alternatively the standard second-order equation of GR with a new dynamical degree of freedom, $f_R$, the equation of motion of which can be obtained by taking the trace of Eq.~(\ref{eq:fr_einstein})
\begin{eqnarray}
\Box f_R = \frac{1}{3}\left(R-f_RR+2f+8\pi G\rho_m\right),
\end{eqnarray}
where $\rho_m$ is the matter density. The new degree of freedom $f_R$ is sometimes dubbed the 
{\it scalaron} in the literature.

Assuming that the background Universe is described by the flat Friedmann-Robertson-Walker (FRW) metric, the line element in the perturbed Universe is written as
\begin{eqnarray}
\rd s^2 = a^2(\eta)\left[(1+2\Phi)\rd\eta^2 - (1-2\Psi)\rd x^i\rd x_i\right],
\end{eqnarray}
in which $\eta$ and $x^i$ are, respectively, the conformal time and comoving coordinates, 
$\Phi(\eta,{\bf x})$ and $\Psi(\eta,{\bf x})$ are the Newtonian potential and perturbation to the 
spatial curvature, and are functions of both time ($\eta$) and space (${\bf x}$); $a$ denotes the scale factor of the Universe where $a=1$ today.

As we are mainly interested in the large-scale structures much smaller than the Hubble scale, and since the time variation of $f_R$ 
is very small in the models considered below, we shall work in the quasi-static limit by neglecting the time derivatives of $f_R$. In this limit, 
the scalaron equation, Eq. \ref{eq:fr_einstein} reduces to
\begin{eqnarray}\label{eq:fr_eqn_static}
\vec{\nabla}^2f_R &=& -\frac{1}{3}a^2\left[R(f_R)-\bar{R} + 8\pi G\left(\rho_m-\bar{\rho}_m\right)\right],
\end{eqnarray}
in which $\vec{\nabla}$ is the three dimensional gradient operator (to be distinguished from the $\nabla$ introduced above), and the overbar represents
 the background value of a quantity. Note that $R$ can be expressed as a function of $f_R$.

Similarly, the Poisson equation which governs the Newtonian potential, $\Phi$, can be simplified to
\begin{eqnarray}\label{eq:poisson_static}
\vec{\nabla}^2\Phi &=& \frac{16\pi G}{3}a^2\left(\rho_m-\bar{\rho}_m\right) + \frac{1}{6}a^2\left[R\left(f_R\right)-\bar{R}\right],
\end{eqnarray}
by neglecting terms involving time derivatives, and using Eq.~(\ref{eq:fr_eqn_static}) to eliminate $\vec{\nabla}^2f_R$.

According to the above equations, there are two potential effects of the scalaron on cosmology: (i) the background 
expansion of the Universe may be modified by the new terms in Eq.~(\ref{eq:fr_einstein}) and (ii) the relationship 
between gravity and the matter density field is modified, which can change the matter clustering and 
growth of density perturbations. Clearly, when $|f_R|\ll1$, we have $R\approx-8\pi G\rho_m$ from Eq.~(\ref{eq:fr_eqn_static}) and 
so Eq.~(\ref{eq:poisson_static}) reduces to the normal Poisson equation in GR; when $|f_R|$ is large, 
we instead have $|R-\bar{R}|\ll8\pi G|\rho_m-\bar{\rho}_m|$ and so Eq.~(\ref{eq:poisson_static}) reduces to 
the normal Poisson equation with $G$ rescaled by $4/3$. 
Note that this factor of 4/3 is the maximum enhancement of
 gravity in $f(R)$ models, independent of the specific functional form of $f(R)$. The choice 
of $f(R)$, however, is important because it governs when and on which scale the enhancement factor 
changes from unity to $4/3$: scales much larger than the range of the modification to Newtonian gravity mediated 
by the scalaron are unaffected and gravity is not enhanced there, while on much smaller scales 
the $4/3$ enhancement is fully realised -- this results in a scale-dependent modification of gravity and therefore a scale-dependent growth rate of structure
 (see Fig. 1).

The relationship between $\Phi$ and $\Psi$ is also changed in $f(R)$ models, with the remaining components of the modified Einstein equation giving
\begin{eqnarray}
\vec{\nabla}^2(\Psi-\Phi) = \vec{\nabla}^2f_R,
\end{eqnarray}
where we have assumed that $|\bar{f}_R|\ll1$. This implies that
\begin{eqnarray}
\vec{\nabla}^2(\Phi+\Psi) = 8\pi G\left(\rho_m-\bar{\rho}_m\right)a^2.
\end{eqnarray}
Therefore the relationship between the lensing potential and the matter density perturbations remains unchanged in $f(R)$ gravity models.

\subsection{The chameleon mechanism}

\label{subsect:fr_cham}

The $f(R)$ gravity would be ruled out by local tests of gravity due to the factor of $4/3$ enhancement to the strength of Newtonian gravity. 
Fortunately, it is well known that, if $f(R)$ is chosen appropriately, the chameleon mechanism \citep{kw2004,ms2007} can 
be exploited to suppress the enhancement allowing this class of models to 
satisfy experimental constraints in high matter density 
regions such as in our Solar system \citep{nv2007,lb2007,hs2007,bbds2008}.

The essence of the chameleon mechanism is as follows. The modifications to the Newtonian gravity can be considered as an 
extra, or fifth force mediated by the scalaron. Because the scalaron itself is massive, the force is of the 
Yukawa type and is suppressed by the exponential factor $ \sim \exp(-mr)$ in which 
$m$ is the scalaron mass and $r$ the distance between two test masses. In high matter density environments, $m$ is very 
heavy and the suppression becomes very strong. In practice, this is equivalent to the 
fact that $|f_R|\ll1$ in high density regions 
because of the exponential suppression, which leads to the GR limit as discussed above.

As a result, the functional form of $f(R)$ is crucial to determine whether 
the fifth force is sufficiently suppressed in high density environments. 
In this work we  study the $f(R)$ model proposed by \citet{hs2007}, for which
\begin{eqnarray}
f(R) = -M^2\frac{c_1\left(-R/M^2\right)^n}{c_2\left(-R/M^2\right)^n+1},
\end{eqnarray}
with $M^2\equiv8\pi G\bar{\rho}_{m0}/3=H_0^2\Omega_m$, where $H$ is the Hubble expansion rate and $\Omega_m$ is the present-day 
fractional density of matter. Hereafter a subscript $0$ always means the current value of 
a quantity.  \citet{hs2007} demonstrated that $|f_{R0}|<0.1$ is required for this model 
to evade Solar system constraints, although the exact value also depends on the behaviour of $f_R$ in the Galaxy.

In the background cosmology, the scalaron $f_R$ always sits at the minimum of the effective potential which governs its dynamics, defined as
\begin{eqnarray}
V_{{\rm eff}}\left(f_R\right) \equiv \frac{1}{3}\left(R-f_RR+2R+8\pi G\rho_m\right),
\end{eqnarray}
around which it oscillates quickly \citep{bdlw2012}. Therefore we have
\begin{eqnarray}
-\bar{R} \approx 8\pi G\bar{\rho}_m-2\bar{f} = 3M^2\left(a^{-3}+\frac{2c_1}{3c_2}\right).
\end{eqnarray}
To match the $\Lambda$CDM background evolution, we need to have
\begin{eqnarray}
\frac{c_1}{c_2} = 6\frac{\Omega_\Lambda}{\Omega_m}
\end{eqnarray}
where $\Omega_\Lambda$ is the current fractional energy density of the dark energy (cosmological constant).

By taking $\Omega_\Lambda\approx0.76$ and $\Omega_m\approx0.24$, we find that $|\bar{R}|\approx41M^2\gg M^2$, and this simplifies the expression of the scalaron to
\begin{eqnarray}
f_R \approx -n\frac{c_1}{c_2^2}\left(\frac{M^2}{-R}\right)^{n+1}.
\end{eqnarray}
Therefore, two free parameters, $n$ and $c_1/c_2^2$, completely specify the $f(R)$ model. Indeed, the latter is related to the value of the scalaron today, $f_{R0}$, as
\begin{eqnarray}
\frac{c_1}{c_2^2} = -\frac{1}{n}\left[3(1+4\frac{\Omega_\Lambda}{\Omega_m})\right]^{n+1}f_{R0}.
\end{eqnarray}
In what follows we shall study three $f(R)$ models with $n=1$ and $|f_{R0}|=10^{-6}, 10^{-5}, 10^{-4}$, which will be referred to as F6, F5 and F4 hereafter, respectively.
These particular parameter choices arise from cluster abundance constraints on the $f(R)$ gravity model. The current constaint found by \citet{svh2009} is
 $|f_{R0}| < 1.3 ^{+1.7}_{-0.6} \times 10^{-4} $ taking into account mass calibration errors.

\begin{figure*}
{\epsfxsize=10.truecm
\epsfbox[77 377 320 717 ]{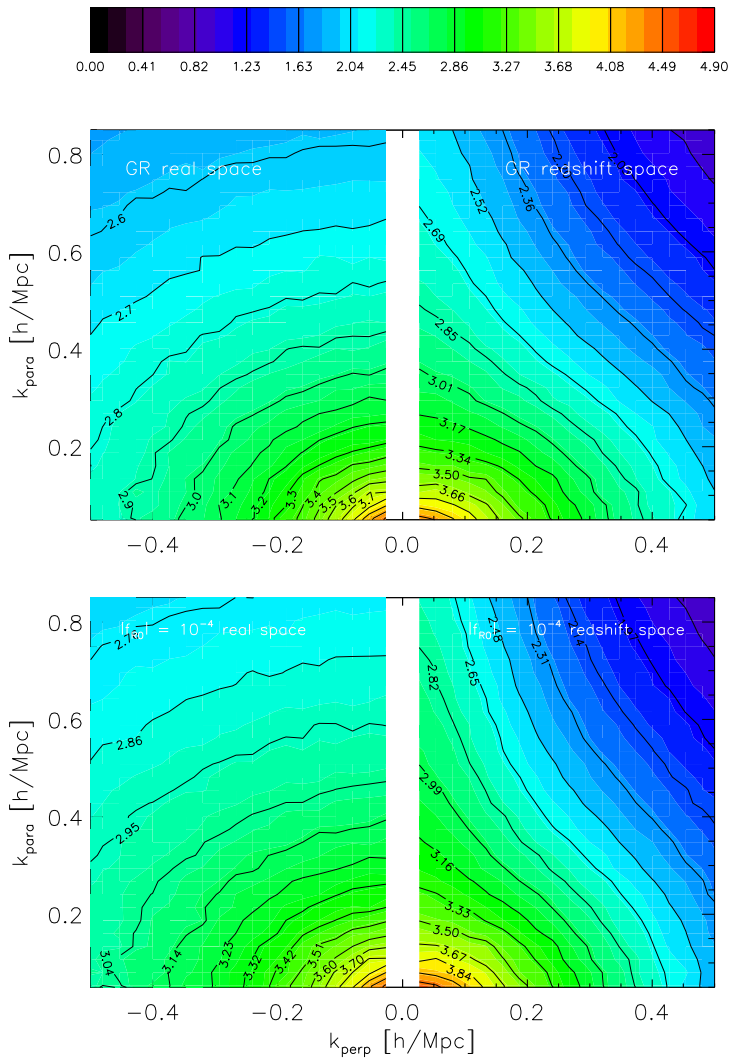}}
\caption{
Top panels: The two dimensional power spectrum measured from the GR simulation as a function of wavenumber perpendicular, $k_{\tiny \mbox{perp}}$,
and parallel, $k_{\tiny \mbox{para}}$, to the line of sight. The colored shading and lines
 represent the amplitude of the power spectrum, $\mbox{log}_{\tiny 10}P$ as indicated by the labels and the scale bar at the top.
The real space power spectrum has been plotted at
 $k_{\tiny \mbox{perp}} \to - k_{\tiny \mbox{perp}}$ to allow comparison with the redshift space $P(k)$.
Bottom panels: The two dimensional power spectrum measured from the F4 simulation as a function of modes perpendicular, $k_{\tiny \mbox{perp}}$,
and parallel, $k_{\tiny \mbox{para}}$, to the line of sight.
\label{2}}
\end{figure*}

\begin{figure*}
{\epsfxsize=10.truecm
\epsfbox[77 377 320 717 ]{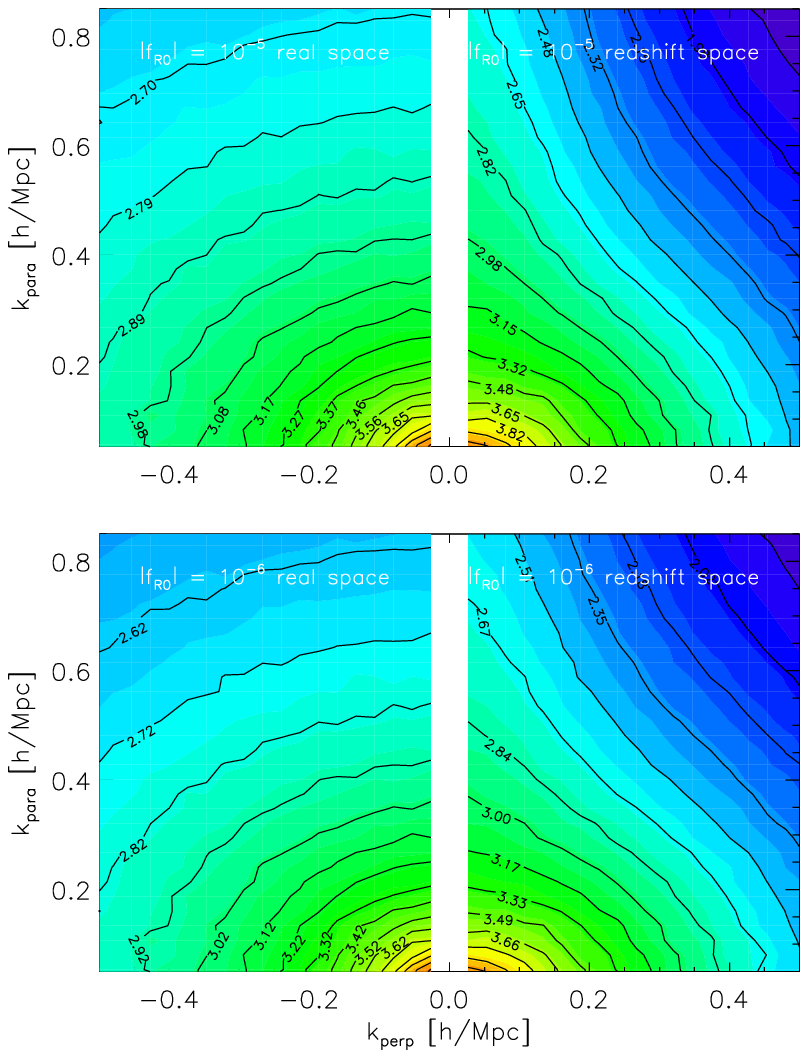}}
\caption{
Top panels: The two dimensional power spectrum measured from the F5 simulation.
Bottom panels: The two dimensional power spectrum measured from the F6 simulation.
The coloured shading correspond to $\mbox{log}_{\tiny 10}P$ as shown in the color bar in Fig. \ref{2}.
As in Fig. \ref{2}, the left panels show the power spectrum in real space and the right panels show redshift space.
\label{3}}
\end{figure*}

\subsection{$N$-body simulations of $f(R)$ gravity}

\label{subsect:fr_nbody}

From Eqs.~(\ref{eq:fr_eqn_static}, \ref{eq:poisson_static}) we can see that, given the matter density field, we can solve for the scalaron field, $f_R$,
 from Eq.~(\ref{eq:fr_eqn_static}) and plug this into the modified Poisson equation (\ref{eq:poisson_static}) to solve for $\Phi$. 
Once $\Phi$ is at hand, we can difference it to calculate the (modified) gravitational force 
which determines how the particles move subsequently. This is exactly what we need to do in $N$-body simulations to follow the evolution of the matter distribution.

The main challenge in the $N$-body simulation of models such as $f(R)$ gravity is to solve the scalaron equation, Eq. \ref{eq:fr_eqn_static}, which is in 
general very nonlinear. For this we need a mesh (or a set of meshes) on which $f_R$ can be solved. This implies that 
mesh-based $N$-body codes are the most suitable for this task. 
On the other hand, particle-based codes are more difficult to apply in this case, as we do not have an analytical formula 
for the modified force law (i.e. the equivalent of $r^{-2}$ in the Newtonian case).

$N$-body simulations for $f(R)$ gravity and related theories have been performed by \citet{oyaizu2008,olh2008,sloh2009,zlk2011,lz2009,lz2010,schmidt2009,lb2011,bbdls2011,dlmw2012}.
However, these simulations were affected by the small box size used and the limited resolution. For this work we have run simulations of 
$f(R)$ cosmologies using the recently developed {\tt ECOSMOG} code \citep{lztk2012}. {\tt ECOSMOG} is a modification of the 
mesh-based $N$-body code {\tt RAMSES} \citep{ramses}, which calculates the gravitational force by first solving 
the Poisson equation on meshes using a relaxation method to get the Newtonian potential and then 
differencing the potential; it does not solve gravity by summing over the forces from nearby 
particles as in done for example in the simulation code
 {\tt GADGET} \citep{springel}. Additional features of the {\tt ECOSMOG} code include:
\begin{enumerate} 
\item The adaptive mesh refinement (AMR), which refines a mesh cell, i.e. splits it into 8 son cells, if the number of particles in a cell 
exceeds a pre-defined number (the refinement criterion). As such it gives higher force resolution in high matter density regions where the chameleon effect is strong and the $f(R)$ 
equation is more nonlinear. The refinement criterion is normally chosen as a particle number between 8 and 12, and in our simulations we adopt a condition of 9 particles. 
This adaptive mesh allows us to reach comparable spatial resolution to codes like {\tt GADGET}.
\item The multigrid relaxation algorithm that ensures quick convergence. The relaxation method finds 
the solution to an elliptical partial differential equation (PDE) on a mesh by iteratively updating 
the initial guess until it converges, i.e., becomes enough close to the true solution. But the rate of 
convergence slows down quickly after the first few iterations. To improve on this, one can 
\lq coarsify\rq  the PDE, i.e., move it to a coarser mesh, solve it there and use the coarse solution to 
improve the solution on the original fine mesh. Unlike other codes, {\tt ECOSMOG} does this on all the AMR meshes, greatly improving the 
convergence properties of the whole code.
\item The massive parallelisation which makes the computation very efficient. This is the key feature 
that enables us to run large simulations such as the ones to be described below, which are beyond 
the reach of any serial code, such as those developed by \citet{lz2009,lz2010,lb2011}.
\end{enumerate}
A convergence criterion is used to determine when the relaxation method has converged. 
In {\tt ECOSMOG} convergence is considered to be achieved when the residual of the 
PDE, i.e., the difference between the two sides of the PDE, is smaller than a predefined 
parameter $\epsilon$. We have checked that for $\epsilon<10^{-8}$ the solution to the PDE no 
longer changes significantly when $\epsilon$ is reduced further. Our choices of $\epsilon$ are listed in Table 1. 
The computational time depends on both the value of $\epsilon$ 
and the model. The $f(R)$ gravity simulations can take  a few times longer to run than the GR simulation.
More details can be found in \citet{lztk2012}.

For the study of redshift space distortions, large simulations boxes are essential to accurately model  behaviour on 
very large scales. For this reason, we have run two sets of simulations, with $L_{\rm box}=1.0h^{-1}$Gpc and $1.5h^{-1}$Gpc respectively. 
The initial
conditions are generated at $z=49$ using the MPgrafic code \citep{mpgrafic}, and 
each suite of F4/F5/F6/GR simulations uses the same initial conditions because at $z=49$ the difference in 
the matter distribution in the different cosmologies is negligible.
The specifications 
of the simulations are summarised in Table~\ref{table:simulations}.

\begin{table*}
\caption{Some technical details of the simulations performed in this work. F6, F5 and F4 are respectively the abbreviations which denote
 the $f(R)$ models with $|f_{R0}|=10^{-6}, 10^{-5}, 10^{-4}$. For all models we have assumed $\Omega_m=0.24$ and $\Omega_{\Lambda}=0.76$, 
and to generate the initial conditions we have used $\sigma_8=0.769$ \citep[in agreement with e.g. ][]{s2009}. We use the same initial conditions 
for all models in each simulation, because at the initial time, $z_i=49$, the difference in the power spectra of different models is 
negligible. $\epsilon$ is the residual for the Gauss-Seidel relaxation used in the code \citep[see][]{lztk2012}, and the two values of the 
convergence criterion are for the coarsest level and finest levels respectively. Other cosmological parameters are a Hubble constant
of $H_{0} = 73 $km/s/Mpc and a scalar spectral index of $n=0.961$.}
\begin{tabular}{@{}lcccccc}
\hline\hline
models & $L_{\rm box}$ & particles & domain meshes & finest meshes & convergence criterion &realisations \\
\hline
$\Lambda$CDM, F6, F5, F4 & $1.0h^{-1}$Gpc & $1024^3$ & $1024^3$ & $65536^3$ & $|\epsilon|<10^{-12}/10^{-8}$ & $1$ \\
$\Lambda$CDM, F6, F5, F4 & $1.5h^{-1}$Gpc & $1024^3$ & $1024^3$ & $65536^3$ & $|\epsilon|<10^{-12}/10^{-8}$ & $6$ \\
\hline
\end{tabular}
\label{table:simulations}
\end{table*}

\section{Models for redshift space distortions \label{RSD}}
In this section we first review the linear perturbation theory of redshift space 
distortions (Section 3.1) before outlining some extended models
which go beyond linear theory (Section 3.2).

\subsection{Linear theory}
Inhomogeneous structure in the universe induces peculiar motions which distort the clustering
pattern measured in redshift space on all scales. This effect must be taken into account when analysing three dimensional datasets which use
redshift to estimate the radial coordinate.
Redshift space effects alter the appearance of the clustering
of matter, and together with nonlinear evolution and bias, lead the measured
power spectrum to depart from  the simple predictions of linear perturbation theory.
The comoving distance to a galaxy, $\vec{s}$,  differs from its true distance, $\vec{x}$, due to its peculiar velocity, $\vec{v}(\vec{x})$
(i.e. an additional velocity to the Hubble flow). The mapping from redshift space to real space is given by
\begin{eqnarray}
\vec{s} = \vec{x} + u_z \hat{z},
\end{eqnarray}
where $u_z = \vec{v}\cdot \hat{z}/(aH)$ and $H(a)$ is the Hubble parameter. This assumes that 
 the distortions take place
 along the line of sight denoted by $\hat{z}$ (N.B. this is the plane parallel approximation). 
This assumption will break down for some pairs of galaxies in a survey which has a wide field of view \citep{Raccanelli}.
Nevertheless, the impact of this systematic on clustering statistics has been shown to be small in comparsion to the effects of nonlinear growth \citep{Samushia}.

On small scales, randomised velocities associated with the motion of galaxies inside virialised structures reduce the power.
The dense central regions of galaxy clusters appear elongated along the line of sight in redshift space, which produces the  \lq fingers
of God\rq\ 
effect in redshift survey cone plots \citep{jackson1972}.
On large scales, coherent bulk flows distort  clustering statistics \citep[see][for a review of redshift space distortions\tc{.}]{hamilton1998}\tc{.}
%{1998ASSL..231..185H}.
For growing perturbations on large scales, the overall effect of redshift space distortions is to enhance the clustering amplitude.
Any difference in the velocity field due to mass flowing from underdense regions to high density regions will alter the volume element, causing
an enhancement of the apparent density contrast in redshift space, $\delta_s(\vec{k})$, compared to that in real space, $\delta_r(\vec{k})$.
This  effect was first analyzed by \citet{kaiser1987} in linear perturbation theory
%\citet{Kaiser:1987qv}
and can be approximated by
\begin{equation}
\label{delta}
\delta_s(k) = \delta_r(k)(1+\mu^2 \beta) ,
\end{equation}
where $\mu$ is the cosine of the angle between the wavevector, $\vec{k}$, and the line of sight, $\beta =  f/b$, $f$ is the linear growth rate
 and  the bias $b=1$ for dark matter.

The \lq Kaiser formula\rq  (Eq. \ref{delta}) relates the overdensity in redshift space to the corresponding value in real space and is the result of
 several approximations, \eg, that the velocity and density perturbations satisfy the linear continuity equation,
\begin{eqnarray}
\delta &=& -f\theta,
\end{eqnarray}
where  $\theta = \vec{\nabla} \cdot \vec{u}$ is the velocity divergence.
All of these assumptions are valid on scales that are well described by linear perturbation theory
and will break down on different scales as the density fluctuations grow \citep[see e.g.][for more details]{jbp2011a}.
As shown in \citet{scoccimarro2004} and \citet{jbp2011a},
%\citet{Scoccimarro:2004tg} and \citet{2011MNRAS.410.2081J},
 the linear regime corresponds to a different range of scales for the matter and velocity fields. 
In particular, linear theory is only a good description of the velocity power spectrum on surprisingly large scales.
We will discuss this further in
Section \ref{Velocity}.

Rather than use the full 2D power spectrum, $P(k,\mu)$, it is common to decompose
the matter power spectrum in redshift space into multipole moments using Legendre polynomials, $L_l(\mu)$, \citep[see e.g.][]{hamilton1998}
\begin{eqnarray}
P(k,\mu) = \sum_{l} P_l(k) L_l(\mu) \, ,
\end{eqnarray}
where the summation is over the order, $l$, of the multipole.
The anisotropy in $P( \vec{k} )$ is symmetric in $\mu$, as $P(k,\mu)=P(k,-\mu)$, so only even values of $l$ are summed over. Each multipole moment is given by
\begin{eqnarray}
P^s_l(k) = \frac{2l+1}{2} \int_{-1}^{1} P(k,\mu) L_l(\mu) \rm{d}\mu \, ,
\end{eqnarray}
where the first two non-zero moments have Legendre polynomials, $L_0(\mu) = 1$ and  $L_2(\mu) = (3\mu^2 - 1)/2$.
Using the linear model in Eq. \ref{delta}, the first three multipole moments are given by
\begin{eqnarray}
\left( {\begin{array}{c}
 P_0(k)   \\
 P_2(k) \\
P_4(k)
 \end{array} } \right)
 &=& P_{\delta \delta}(k)
\left( {\begin{array}{c}
1 + \frac{2}{3}\beta + \frac{1}{5}\beta^2   \\
\frac{4}{3}\beta + \frac{4}{7}\beta^2 \\
\frac{8}{35}\beta^2
 \end{array} } \right) \, ,
\label{moments}
\end{eqnarray}
where $P_{\delta \delta}(k) = \langle|\delta(k)|^2\rangle$ denotes the real space matter power spectrum. Note we have neglected the superscript $s$ here
for clarity. 
In practice, $P_{\delta \delta}(k)$ cannot be obtained directly for a real survey without making approximations \citep[\eg][]{be1994}.%{1994MNRAS.270..183B}.

In this paper we consider the estimator for $\beta$ suggested by \citet{cfw1994},
%\citet{Cole:1993kh},
which is the ratio of the quadrupole to monopole moments  of the redshift
space power spectrum, $P_2(k)/P_0(k)$, which is independent of the real space power spectrum.

\subsection{Quasi-linear and Non-linear models for the redshift space power spectrum \label{rsdmodels}}

Assuming the line of sight component of the peculiar velocity 
is along the $z$-axis, the  full nonlinear relation between the real and redshift space power spectrum can be written as \citep{scf1999}
%\citep{Scoccimarro:1999ed}
\begin{eqnarray}
\label{NL}
P^s(k,\mu) =  \int \frac{\rm{d}^3 \bf{r}}{(2\pi)^3} e^{-i \bf{k} \cdot \bf{r}} \langle e^{i\lambda \Delta u_z} [\delta({\bf{x}}) -  \theta({\bf{x}})] \nonumber  \\
 \times [\delta({\bf{x'}}) - \theta({\bf{x'}})]\rangle \, ,
\end{eqnarray}
where $\lambda =  k \mu$, $u_z$ is the comoving peculiar velocity
along the line of sight, $\Delta u_z = u_z({\bf{x}}) - u_z(\bf{x'})$,  $\bf{r} = \bf{x} -\bf{x'}$, $\theta = \nabla_z \cdot u_z$,
 and the only approximation made is the plane parallel approximation.
At small scales (as $k$ increases) the exponential component damps the power, representing the impact of randomised velocities inside gravitationally
bound structures.

\begin{figure*}
{\epsfxsize=12.5truecm
\epsfbox[74 373 545 616]{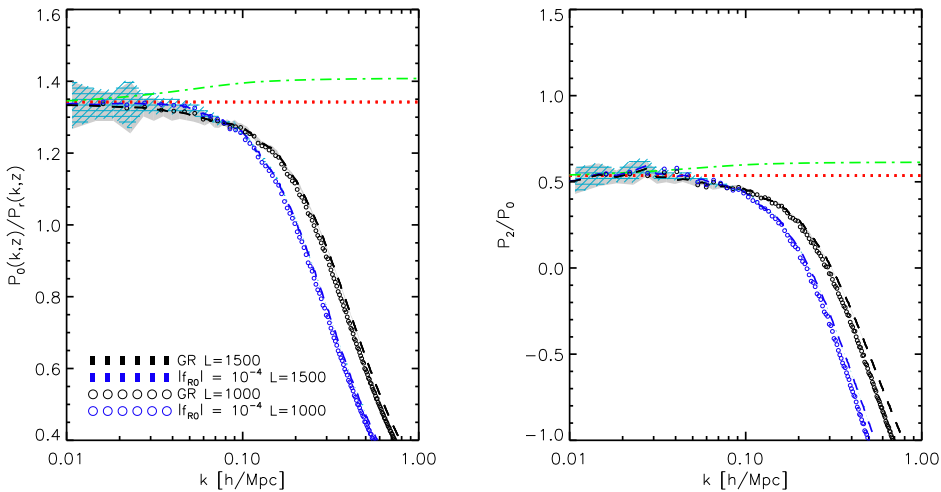}}
\caption{
Left panel: The ratio of the monopole of the redshift space power spectrum to the real space power spectrum at $z=0$ in the
F4 (blue) and GR (black) cosmologies. We plot the  ratio measured from two simulation boxes: $L_{\tiny \mbox{box}} = 1000$ Mpc$/h$ (circles)
and $L_{\tiny \mbox{box}} = 1500$ Mpc$/h$ (dashed lines). The linear theory prediction 
for each model is shown as a green dot dashed line for the $f(R)$ model
and a red dashed line for GR. The shaded regions represent the errors on the ratios measured from six realisations of the $f(R)$ (blue hatched) and GR (grey solid)
cosmologies.
Right panel: The ratio of the quadrupole to monopole moment of the redshift space
$P(k)$ at $z=0$ in the
F4 (blue) and GR (black) cosmologies.
\label{p2p0f4}}
\end{figure*}

\begin{figure*}
{\epsfxsize=12.5truecm
\epsfbox[70 368 550 617]{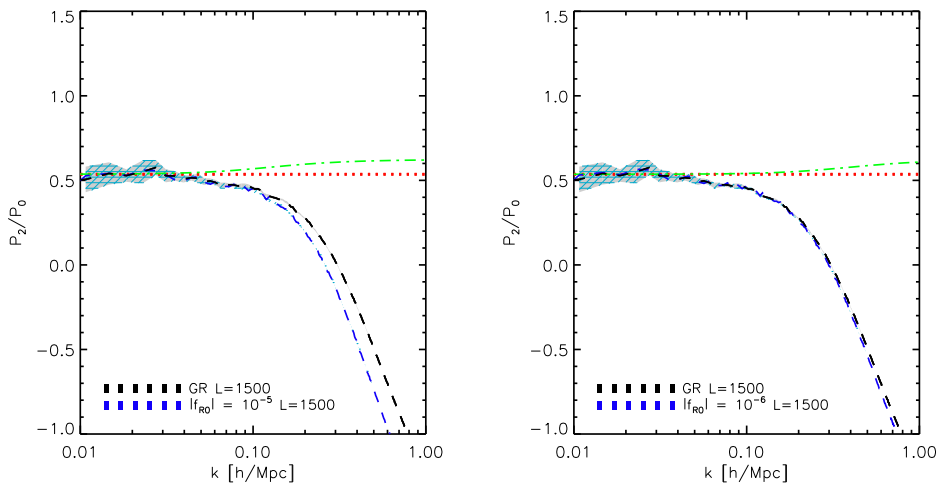}}
\caption{
Left panel: The ratio of the quadrupole to monopole moment of the redshift space power spectrum  at $z=0$ in the
F5 (blue) and GR (black) cosmologies.
Right panel: The same ratio of the quadrupole to monopole moment of the redshift space power spectrum  at $z=0$ in the
F6 (blue) and GR (black) cosmologies. 
The linear theory prediction for each model is shown as a green dot dashed line for the $f(R)$ model
and a red dashed line for GR. The shaded regions represent the errors on the ratios measured from six simulations of the $f(R)$ (blue hatched) and GR (grey solid)
cosmologies. The simulation box size used was $L_{{\rm box}} = 1500$Mpc$/h$ on a side.
\label{p2p0f5f6}}
\end{figure*}

Simplified models for redshift space distortions are frequently used.
Examples include multiplying Eq. \ref{moments} by a factor which
attempts to take into account small scale effects, invoking either a
Gaussian or exponential distribution
of peculiar velocities \citep{pd1994}. %\citep{Peacock:1993xg}.
A popular phenomenological example of this which incorporates the damping effect of velocity dispersion on small scales is the so-called \lq dispersion model\rq \ \citep{pd1994},
%\citep{Peacock:1993xg},
\begin{eqnarray}
P^s(k,\mu) =  P^r(k) (1+\beta \mu^2)^2 \frac{1}{(1 + k^2 \mu^2 \sigma_p^2/2)} \, ,
\end{eqnarray}
where $\sigma_p$ is the pairwise velocity dispersion along the line of sight, which is treated as a parameter to be fitted to the data.

The linear model for the redshift space power spectrum can be extended by keeping
the nonlinear velocity power spectra terms in Eq. \ref{NL}.
The velocity divergence auto power spectrum is the ensemble average, $P_{\theta \theta} = \langle |\theta|^2 \rangle $
 where $\theta = \vec{\nabla}\cdot \vec{u}$ is the velocity
divergence. The cross power spectrum of the velocity divergence and matter density is $P_{\delta \theta}= \langle |\delta \theta| \rangle$.
%\citet{Scoccimarro:2004tg}
\citet{scoccimarro2004} proposed the following model for the redshift space power spectrum in terms of $P_{\delta \delta}$, the nonlinear
matter power spectrum,
 $P_{\theta \theta}$ and $P_{\delta \theta}$,
\begin{eqnarray}
\label{SM}
&&P^s(k,\mu)=\\  \nonumber
&&  \left( P_{\delta \delta}(k) + 2 \mu^2 P_{\delta \theta}(k) + \mu^4P_{\theta \theta}(k)\right)
\times e^{-( k \mu \sigma_v )^2} \, ,
\end{eqnarray}
where   $\sigma_v$ is the 1D linear velocity dispersion given by
\begin{eqnarray}
 \sigma^2_v = \frac{1}{3}\int\frac{P_{\theta \theta}(k)}{k^2} {\rm d}^3k.
\end{eqnarray}
In linear theory, $P_{\theta \theta}$ and $P_{\delta \theta}$ take the same form as $P_{\delta \delta}$ and depart from this at different scales.
Using  a simulation  with 512$^3$ particles in a box of length $479 h^{-1}$Mpc \citep{ysd2001}, %\citep{Yoshida:2001vf}, \citet{Scoccimarro:2004tg}
\citet{scoccimarro2004} showed that this simple ansatz  for $P_s(k,\mu)$ was an improvement over the Kaiser formula when comparing to 
the results of $N$-body simulations in a $\Lambda$CDM
cosmology.

In  nonlinear models for the power spectrum in redshift space there is a degeneracy between the
nonlinear bias, the difference between the clustering of dark matter and halos or galaxies, and the
scale dependent damping due to velocity distortions on small scales.
This degeneracy will complicate any measurement of the growth rate using redshift space clustering information on small scales.
In addition these models assume that the growth rate is scale independent as is the case in general relativity.
As shown in  Fig. \ref{growthrate}, the $f(R)$ model has scale dependent growth rates which would have to be included when
fitting any model to the measured redshift space power spectrum.

In this paper we analyze the redshift space clustering of the dark matter in $f(R)$ and $\Lambda$CDM cosmologies.
We restrict our analysis to the linear and quasi-linear regime where the bias is typically assumed to be scale independent 
and so our results can be more easily extended
 to linearly biased tracers of the dark matter field \citep[but see][for counterexamples]{a2008}.
We will model the redshift space distortions in the clustering of halos in both of these cosmologies in future work.

We shall use the following model for the 2D redshift space power spectrum,
\begin{eqnarray}
\label{linearmodel}
P(k,\mu) &=& P_{\delta \delta}(k) +  2\mu^2P_{\delta \theta}(k) + \mu^4P_{\theta \theta}(k) \, ,
\end{eqnarray}
where the first two multipole moments are given by
\begin{eqnarray}
\left( {\begin{array}{c}
 P_0(k)   \\
 P_2(k)
 \end{array} } \right)
 &=&
\left( {\begin{array}{ccc}
1 & \frac{2}{3} & \frac{1}{5}   \\
0 & \frac{4}{3} & \frac{4}{7}
 \end{array} } \right) \,
 \left(  {\begin{array}{c}
P_{\delta \delta}(k) \\
P_{\delta \theta}(k) \\
P_{\theta \theta}(k)
 \end{array} } \right) \, .
\label{linearmodelmoments}
\end{eqnarray}
This model has been shown to be a good fit to the power spectrum in
redshift space measured from simulations at $z<1$ \citep{jbp2011a,jbp2011b}.
%\citep{2011MNRAS.410.2081J, 2011ApJ...727L...9J}.

\section{Results \label{RESULTS}}

In Section~\ref{section_pk} we present measurements of the  redshift space power 
spectra for both general relativity and the  $f(R)$ models.
In Section~\ref{Velocity} we present the velocity power spectrum measured from the simulations.
We attempt to extract both the matter and velocity power spectra from the
two dimensional redshift space power spectrum 
using a quasi-linear model for the redshift space $P(k)$ 
 in Section~\ref{reconstructing}. In Section~\ref{moments}
we examine how well the moments of the redshift space power spectrum can be recovered
using this quasi linear model.

\subsection{The power spectrum in redshift space \label{section_pk}}

In Figs.~\ref{2} and \ref{3} we plot the two dimensional power spectrum measured at $z=0$ from the GR and F4 simulations  and the F5 and F6 
simulations respectively
 as a function of wavenumber perpendicular, $k_{\tiny \mbox{perp}}$,
and parallel, $k_{\tiny \mbox{para}}$, to the line of sight. The color contours and lines represent 
the amplitude of the power spectrum, $\mbox{log}_{10}P$. In each figure the real space power 
spectrum has been plotted as $k_{\tiny \mbox{perp}} \to - k_{\tiny \mbox{perp}}$ in order to 
allow a side-by-side comparison  with the redshift space $P(k)$. These figures clearly show that the 
spherical symmetry seen in the real space power spectrum (left panel) is distorted in 
redshift space (right panel): on large scales ($k\rightarrow 0$),  the amplitude of the 
redshift space power spectrum is increased compared to that in real space, whereas on small scales,  
the power spectrum is damped and elongated along the line of sight in redshift space compared to real space. 
These effects 
were first convincingly
observed in the clustering of galaxies measured by the 2dFGRS \citep{p2001}
and again recently by
the SDSS-III BOSS survey and the WiggleZ Dark Energy Survey, c.f.~Fig.~3 in \citet{retal2012} and Fig.~2 in \citet{betal2011}.
%\citet{2011MNRAS.415.2876B}.
These two effects are more pronounced in the F4 simulation where the large scale boost and small scale damping appear larger than in GR.
Overall, the redshift space $P(k)$ for the $f(R)$ model appears far more distorted and asymmetrical than the corresponding redshift space $P(k)$ in GR.

In the left panel of Fig.~\ref{p2p0f4} we plot the ratio of the monopole of the redshift space power spectrum to the real space power spectrum at $z=0$ measured from the
GR (black) and F4 (blue) simulations. The right panel of this figure shows the ratio of the quadrupole to monopole moment of the redshift space
power spectrum for both models at $z=0$. The redshift space power spectra are obtained from the simulations after averaging over the
$P(k)$ obtained by treating the $x, y$ and $z$ directions  in turn as the lines of sight. The errors on the ratios are plotted as a
blue hatched region for the $f(R)$ cosmology and as a solid grey region for GR and represent the scatter amongst six realisations.
Note we have compared the errors obtained from six simulations 
to those from ten simulations from \citet{jbp2011b}, which
 have the same box size and 
particle number, and find a 20\% decrease in the $z=0$ error on the largest scale, $k \sim 0.01 h/$Mpc, when we use six simulations instead of ten.
The linear theory predictions, given by Eq.~\ref{moments}, are shown as a green dot dashed line for the $f(R)$ model
and a red dashed line for GR. These predictions use the linear growth rate for each model which for GR is $f(z=0) = 0.42$ and is a scale dependent factor, $f(k)$,
for the $f(R)$ cosmology (see Fig. 1).

It is clear from Fig.~\ref{p2p0f4} that the redshift space power spectrum in the $f(R)$ model has a different shape compared to that in GR. Firstly,
there is an increased boost in the clustering signal on large scales, $k < 0.07 h/$Mpc, due to increased bulk flows into
overdense regions seen in the modified gravity simulation. 
Nevertheless, the measurement from the simulation is significantly below the linear theory prediction for this cosmology.
The small scale damping due to incoherent motions within virialised structures is also more pronounced in the $f(R)$
 cosmology compared to GR on scales $k> 0.1h/$Mpc.
It is clear from this plot that the linear perturbation theory limit is only attained on extremely large scales $(k <0.02 h/$Mpc$^{-1}$$)$ for each model.
On these scales the two models cannot be distinguished within the error bars. 
The large scale boost in the redshift space power spectrum in the $f(R)$ cosmology compared to GR is less pronounced in the
quadrupole to monopole moment ratio plotted in the right panel of Fig. \ref{p2p0f4}. The increased damping of the redshift space power spectrum found
in the modified gravity model can again be seen on small scales ($k>0.1 h/$Mpc). 
The increased damping of the redshift space power spectrum on small scales is a generic feature of  $f(R)$ gravity. 
This is because the fifth force causes the particles to move faster, \ie, with a larger velocity dispersion than they do in GR for the same mass structure.
%In general, the redshift-space power spectrum gets more damped on small scales in the $f(R)$ gravity because the fifth force causes the particles to move faster, \tc{\ie} with a larger velocity dispersion, than they do in GR.

In Fig.~\ref{p2p0f4} we also compare the measured ratios from two different simulation boxes of  $L_{\tiny \mbox{box}} = 1000 $ Mpc$/h$ (circles) and 
$L_{\tiny \mbox{box}} = 1500 $ Mpc$/h$ (dashed lines) on a side. Each simulation has the same number 
of particles, $1024^3$, but with different resolutions. In the  $L_{\tiny \mbox{box}} = 1500 $ Mpc$/h$ simulation, the 
domain grid\footnote{Here the domain grid is the finest grid that is uniform across the computational box. }
  has $1024^3$ cells with a refinement 
criterion equal to 9 particles (which means that a cell is refined into eight \lq son\rq \, cells if the number of particles inside it exceeds 9).  
The $L_{\tiny \mbox{box}} = 1000 $ Mpc$/h$ simulation was run with the same domain grid and  refinement criterion. 
As a result, both the mass and force resolutions are 
higher in the  $L_{\tiny \mbox{box}} = 1000 $ Mpc$/h$ simulation. 
The force resolution is 15.3 kpc$/h$ and 22.9 kpc$/h$ in the 1 Gpc$/h$ and 1.5 Gpc$/h$ box simulations respectively. 
This can be
compared to the force resolution of 781 kpc$/h$ for
the 400 Mpc$/h$ box and 125 kpc$/h$ for the 64 Mpc$/h$
box used by \citet{svh2009}.
In Fig.~\ref{p2p0f4} the higher resolution simulation shows more damping on small scales, and 
this is clearly
seen in the quadrupole to monopole moment plotted in the right panel. This difference in the results from different computational boxes 
 shows that large volume high resolution simulations are essential in order to
accurately resolve the velocity field on scales $k >0.3 h/$Mpc and provide accurate predictions for the power spectrum in redshift space.
In this paper we restrict our study to scales $k<0.3 h/$Mpc where the velocity field is accurately resolved in both simulations. We will only show
 results from the $L_{\tiny \mbox{box}} = 1500$ Mpc$/h$ simulation in the rest of the paper unless otherwise stated.

In Fig.~\ref{p2p0f5f6} we plot the ratio of the quadrupole to monopole moment of the redshift space power spectrum at $z=0$ measured in the F5 
(left panel) and F6 simulations (right panel) along with the measured ratios in GR. 
The fifth force is more strongly suppressed in both of these models compared to the F4 model. As a 
result we find no detectable boost in the power spectrum on large scales compared to GR and less damping on small scales compared to F4.
The F5 model in the left panel of Fig.~\ref{p2p0f5f6} still shows significantly more damping on 
small scales compared to GR, whereas the difference between the F6 model and GR at $z=0$ is very small 
even down to small scales ($k\sim0.3h/$Mpc). This implies that the particles inside halos are 
not significantly affected by the fifth force in the F6 model, while in F5 they have started to feel an 
effect.

\begin{figure}
{\epsfxsize=7.5truecm
\epsfbox[68 368 266 704]{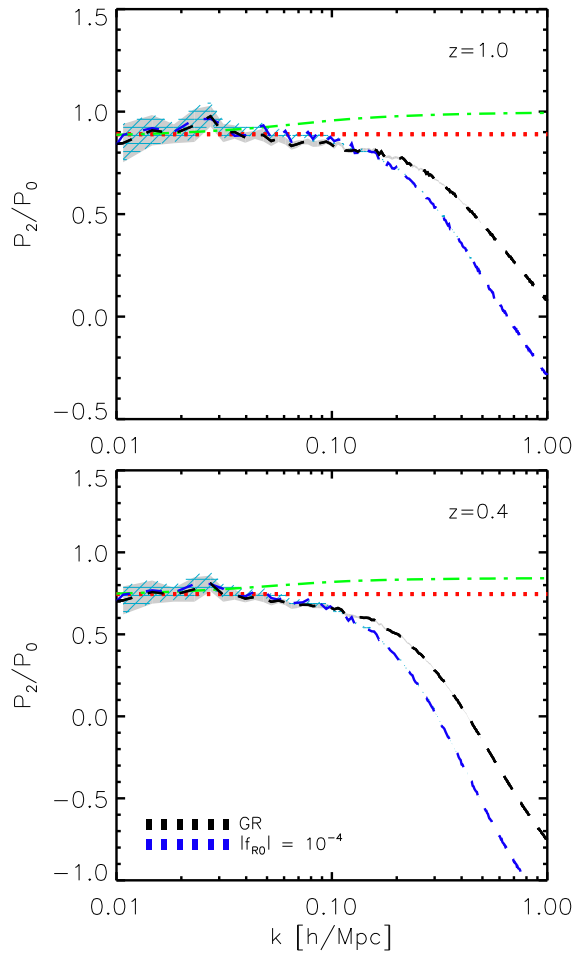}}
\caption{
The ratio of the quadrupole to  monopole moment of the redshift space power spectrum
for $\Lambda$CDM and the F4 model at $z=0.4$ (bottom panel) and $z=1$ (top panel).
The linear theory prediction at each redshift is shown as a green dot dashed line for the $f(R)$ model
and a red dashed line for GR.
\label{p2p0f4z1}}
\end{figure}

Fig. \ref{p2p0f4z1} shows how the quadrupole to monopole moment ratios of the redshift space power spectrum
for GR and the F4 model change with redshift. The lower (upper) panel shows the measured ratios at $z=0.4$ ($z=1.0$), 
together with the linear theory predictions for each model (red dotted line for GR and green dot-dashed line for F4) 
at the same redshift. We can clearly see that the linear theory predictions agree with the measured ratios on 
slightly smaller scales (down to $k\sim0.05 h/$Mpc) at $z=1$ than $z=0$ for both models, as expected.  
The ratio $P_2/P_0$ in the F4 models is slightly larger than GR on large scales but it suffers stronger 
damping on small scales compared to GR, such that the ratio becomes smaller than GR on nonlinear scales. This 
large scale boost of the power spectrum and extra small scale damping compared to GR are more pronounced at 
higher redshifts, which seems to contradict the naive expectation given that the fifth force is weaker then. 
This is because the fifth force effect has already been felt well before $z=1$ as the screening is weaker at these epochs, 
after which the GR result 
slightly catches up and so the difference from F4 is reduced. Indeed, the same trend can also be seen in the linear perturbation results plotted in this figure.

\begin{figure}
{\epsfxsize=6.truecm
\epsfbox[82 378 207 696]{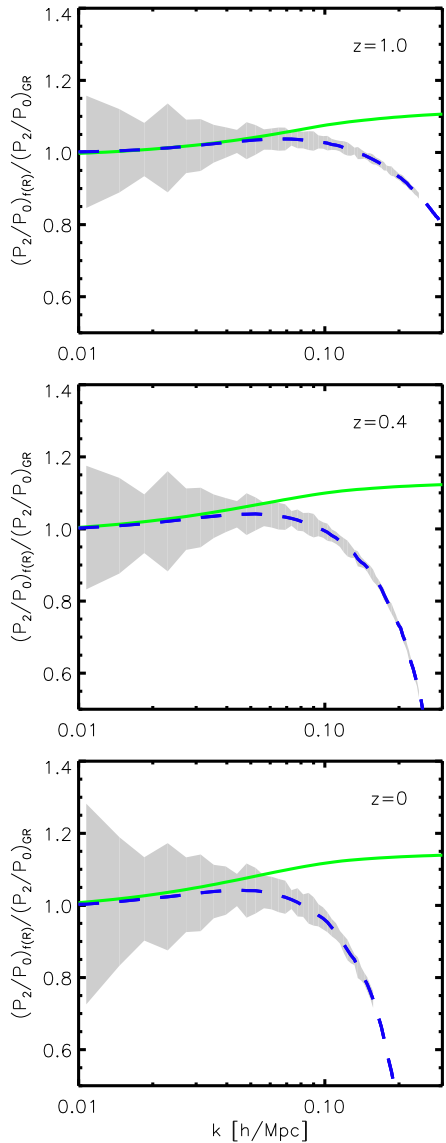}}
\caption{
The ratio of $P_2/P_0$ in the F4 model to that in GR at $z=0$ (bottom panel), $z=0.4$ (middle panel) and $z=1$ (top panel) measured 
from the simulations. 
The linear theory prediction for this ratio at each redshift is shown as a solid green line.
\label{new}}
\end{figure}

 Fig.~\ref{new} shows the ratio of $P_2/P_0$ in the F4 models to that in GR at $z=0$, $z=0.4$ and $z=1$ 
measured from the simulations, together with the linear theory predictions. Linear theory predicts that the relative difference of $P_2/P_0$ between F4 and GR is larger at lower redshifts as expected. However, the relative difference measured in simulations on large scales is smaller at lower redshifts. This is due to the extra damping in the F4 models. This damping becomes stronger at lower redshifts and overcomes the linear enhancement of $P_2/P_0$ in the F4 models.

The strong enhancement in the small scale  damping in the F4 and F5 models
compared to that in GR, which we have seen above, could be a clear signal of modified gravity that persists at higher redshifts. 
Of course, here we are only talking about the dark matter power spectrum and, as 
discussed in Section~\ref{rsdmodels}, models for the redshift space power 
spectrum on small scales need to account for nonlinear bias effects. As this is most relevant 
when studying the clustering of halos in redshift space we leave this analysis
to future work.

\subsection{The velocity power spectrum \label{Velocity}}

\begin{figure*}
{\epsfxsize=13.truecm
\epsfbox[79 371 536 685]{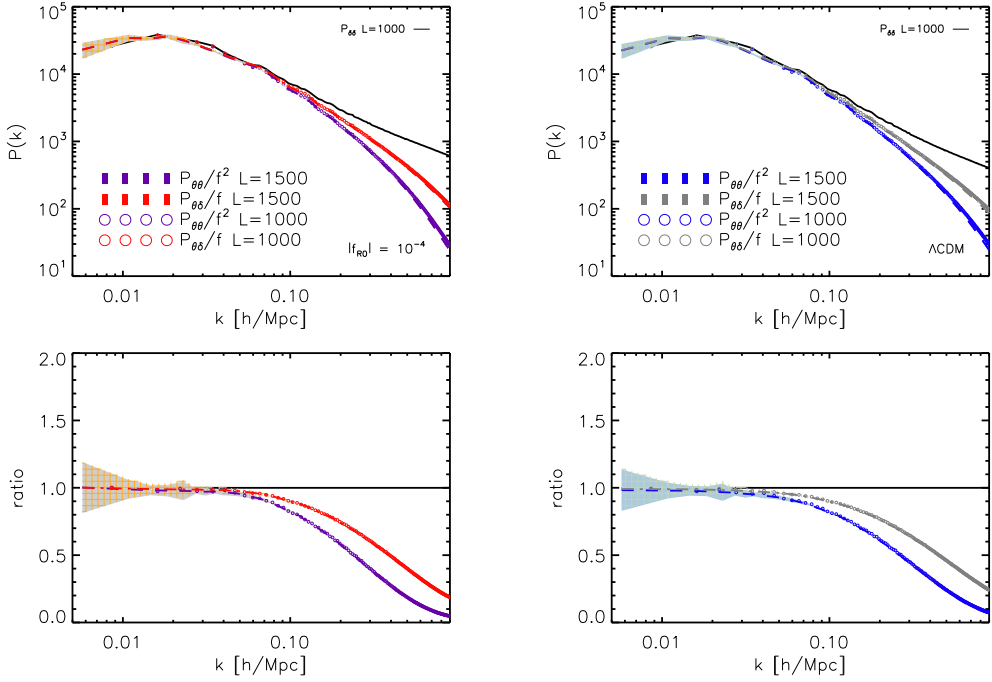}}
\caption{
Left panel: The velocity divergence auto, $P_{\theta \theta}/f^2$ (purple) and cross, $P_{\delta \theta}/f$ (red)
power spectrum at $z=0$ measured from the F4 model simulation. The lower ratio plot shows
$P_{\theta \theta}/P_{\delta \delta}f^2$ (purple) and $P_{\delta \theta}/P_{\delta \delta}f$ (red).
Right panel: The velocity divergence auto, $P_{\theta \theta}/f^2$ (blue) and cross, $P_{\delta \theta}/f$ (grey)
power spectrum at $z=0$ measured from the $\Lambda$CDM simulation. The lower ratio plot shows
$P_{\theta \theta}/P_{\delta \delta}f^2$ (blue) and $P_{\delta \theta}/P_{\delta \delta}f$ (grey).
The nonlinear matter power spectrum is plotted as a solid black line for each model. The measured $P(k)$ from the
$L_{\tiny \mbox{box}} = 1500 $ Mpc$/h$  and  $L_{\tiny \mbox{box}} = 1000 $ Mpc$/h$  simulations are plotted as dashed lines and
circles respectively.
\label{pddptt}}
\end{figure*}

The nonlinear evolution of velocity fields on large scales can have a significant impact
on the redshift space power spectrum. %Scoccimarro (2004)
\citet{scoccimarro2004} showed that the velocity field is more sensitive to tidal gravitational fields
compared to the density field on large scales. Taking these nonlinear velocity effects into account results in an
improved model for the power spectrum in redshift space in the quasi-linear regime \citep{jbp2011a}. %\citep{2011MNRAS.410.2081J}.

Measuring the velocity power spectrum from simulations can be difficult. The method suggested by \citet{scoccimarro2004}
%\citet{Scoccimarro:2004tg}
allows a mass weighted velocity field to be constructed but is limited by the fact that it is the momentum field
which is calculated on a grid and so the velocity field in empty cells is artificially set to zero \citep{ps2009}. %\citep{2009PhRvD..80d3504P}.
Another limitation of this method is that most calculations require the volume weighted velocity field instead of the mass weighted field. 
Using a Delaunay tessellation
of a discrete set of points allows the desired volume weighted velocity field to be constructed accurately on small scales.
We use the publicly available DTFE code \citep{sv2000, vis2009,cv2011}
%\citep{2011arXiv1105.0370C, 2000A&A...363L..29S}
to construct the velocity divergence field directly.
This code  constructs the Delaunay tessellation from a discrete set of points and interpolates the field values onto a user defined grid.
For the $L_{\tiny \mbox{box}} = 1500 $ Mpc$/h$ simulation
we are able to generate the velocity auto, $P_{\theta \theta}$, and cross power spectrum, $P_{\delta \theta}$, on a $1024^3$ grid.
The density field is interpolated onto the 
grid using the cloud-in-cell scheme. The resolution of the mesh means that 
mass assignment effects are negligible on the scales of interest here.

Fig.~\ref{pddptt} shows the $z=0$ nonlinear velocity and matter power spectra measured from the F4 (left panels) and GR (right panels) simulations.
The errors calculated from the scatter amongst six simulations
are shown as a hatched region for the cross power spectrum, $P_{\delta \theta}$, and as a solid shaded region
for the auto power spectrum,$P_{\theta \theta}$.
We show the velocity power spectrum from both the $L_{\tiny \mbox{box}} = 1500 $Mpc$/h$ (dashed lines) and $L_{\tiny \mbox{box}} = 1000 $Mpc$/h$ (circles) simulations.
The lower panel in each case shows the ratio of $P_{\theta \theta}/P_{\delta \delta}f^2$ and $P_{\delta \theta}/P_{\delta \delta}f$ for each model. 
 The scales where
the linear continuity equation breaks down are shown by the departure of the measured ratios
 from unity. This occurs on slightly larger scales for GR than for the $f(R)$ model.
The fact that the velocity power spectrum departs from linear theory on larger scales than the density field agrees
 with what has been noted previously by \citet{scoccimarro2004} and \citet{jbp2011a}.
%\citet{Scoccimarro:2004tg, 2011MNRAS.410.2081J}.
We find that $P_{\delta \delta}$ and $P_{\theta \theta}$ differ by up ~20\% at
$k < 0.1 h/$Mpc.

\begin{figure*}
{\epsfxsize=13.5truecm
\epsfbox[68 371 365 697]{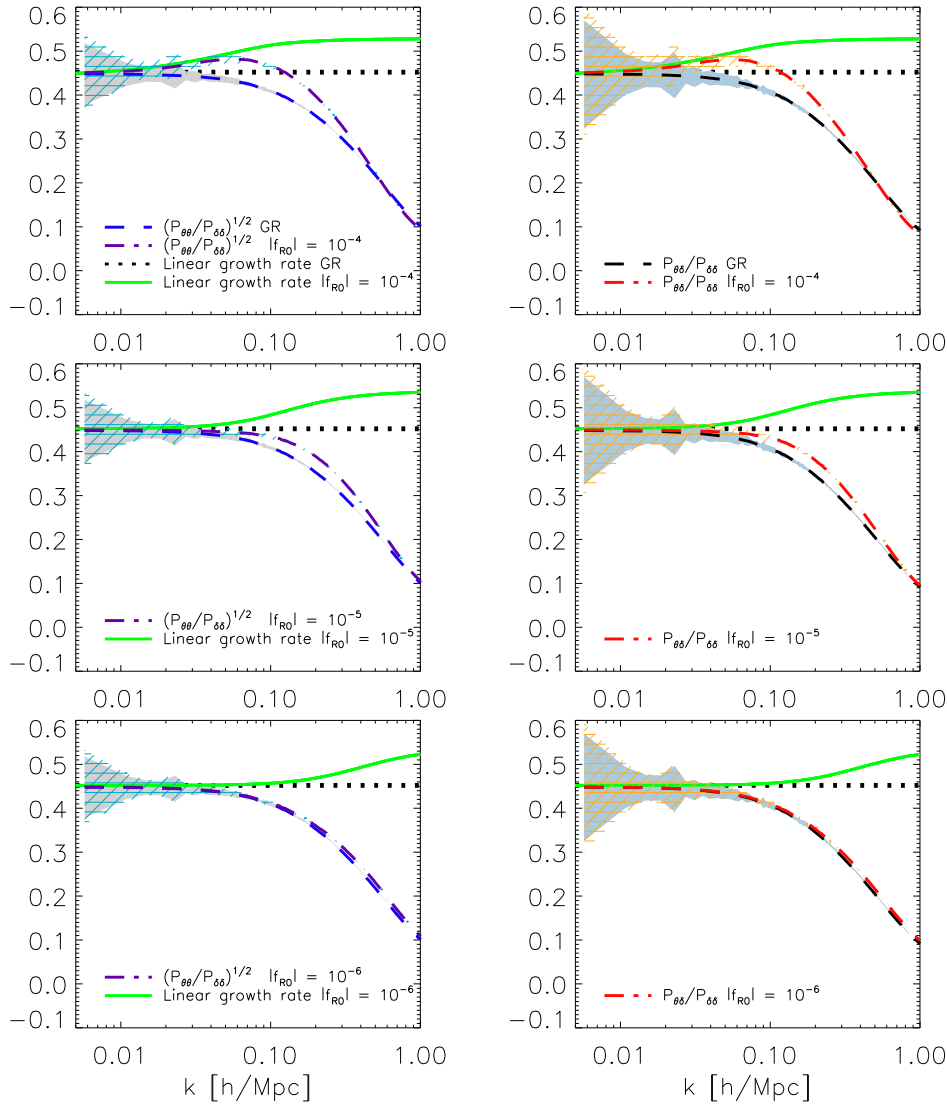}}
\caption{
Top left panel: The ratio
$\sqrt{P_{\theta \theta}/P_{\delta \delta}}$ at $z=0$ for the  F4 (dashed purple) and $\Lambda$CDM (dashed blue) models.
The same ratio is shown for F5 and F6 models in the middle and bottom left hand panels respectively.
Top right panel: The ratio $P_{\delta \theta}/P_{\delta \delta}$  at $z=0$ for the F4 (dashed red) and $\Lambda$CDM (dashed black) models.
The same ratio is shown for F5 and F6 in the middle and bottom right hand panels respectively.
In all panels
the variable  $\theta  = \vec{\nabla} \cdot \vec{u}$. The errors for the $f(R)$ model are shown as hatched shaded regions
and as solid shaded regions for $\Lambda$CDM. The linear growth rate is plotted as a solid green line and dashed black line for each
$f(R)$ model and $\Lambda$CDM respectively.
\label{f4theta}}
\end{figure*}

When the velocity power spectrum is normalised using the linear growth rate as in Fig. \ref{pddptt},
the ratio of the nonlinear velocity and matter power spectra look very similar in both the
standard and the modified gravity cosmologies. If instead we choose not to normalise $\theta$ using the growth rate, we get the curves shown in
the top row of Fig. \ref{f4theta} for the ratio $\sqrt{P_{\theta \theta}/P_{\delta \delta}}  $  (left panel) and
$P_{\delta \theta}/P_{\delta \delta}  $ (right panel) for the F4 and $\Lambda$CDM cosmologies.
Note in
linear perturbation theory these two ratios equal the
linear growth rate, $f$ , which is plotted as a solid
green line for the $f(R)$ model and as a dotted black
line for $\Lambda$CDM in Fig. 9.
There is clearly a large difference in the amplitude and shape of these ratios in the two models on scales $k>0.03 h/$Mpc.
 It is interesting that the ratios agree with the predictions for the linear growth rate on  scales where
 the two models can be distinguished within the errors, shown by the 
hatched shaded region for $f(R)$ and solid shaded region for $\Lambda$CDM. This is in contrast to 
the multipole moments of the redshift space power spectrum, Fig.~\ref{p2p0f4}, where
the Kaiser model predictions using the linear growth rate only agree with the measured $P(k)$ on extremely large
scales where the two cosmologies could not be distinguished within the errors. The linear growth rate for the F4 model
differs from that in $\Lambda$CDM by up to 20\% for $k < 0.1 h/$Mpc (see Fig. 1).
\begin{figure}
{\epsfxsize=7.5truecm
\epsfbox[77 379 269 703]{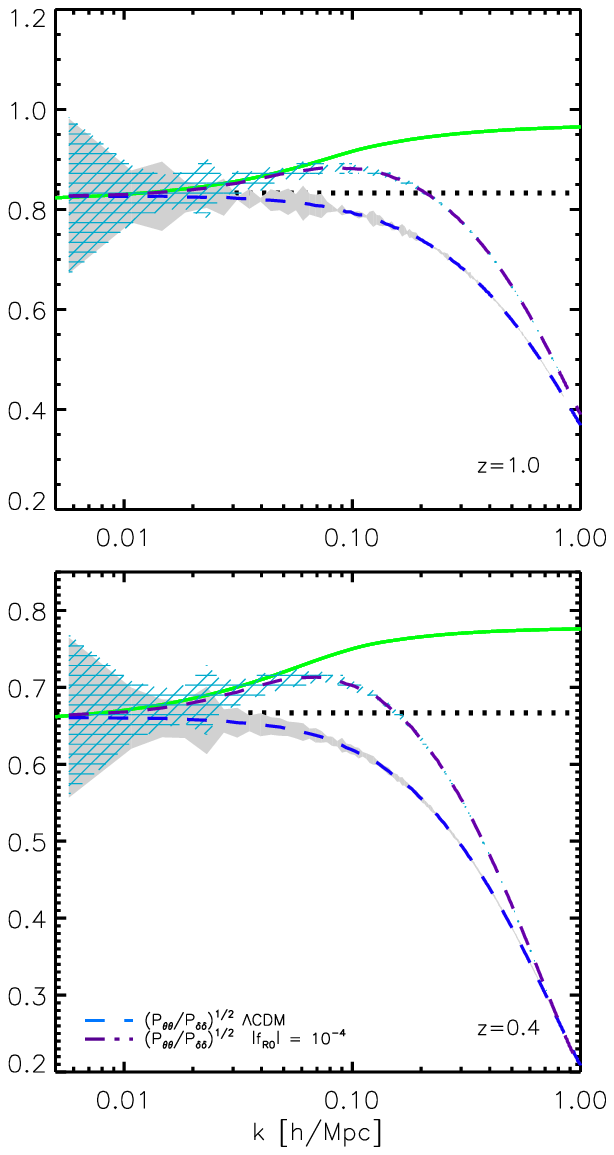}}
\caption{
The ratio
$\sqrt{P_{\theta \theta}/P_{\delta \delta}}$ at $z=0.4$ (lower panel) and $z=1$ (upper panel)
for the F4 (dashed purple) and $\Lambda$CDM (dashed blue) models.
The linear growth rate at each redshift is plotted as a solid green line and dashed black line for the
$f(R)$ model and $\Lambda$CDM respectively.
\label{f4thetaz1}}
\end{figure}

\begin{figure}
{\epsfxsize=6.truecm
\epsfbox[80 373 214 695]{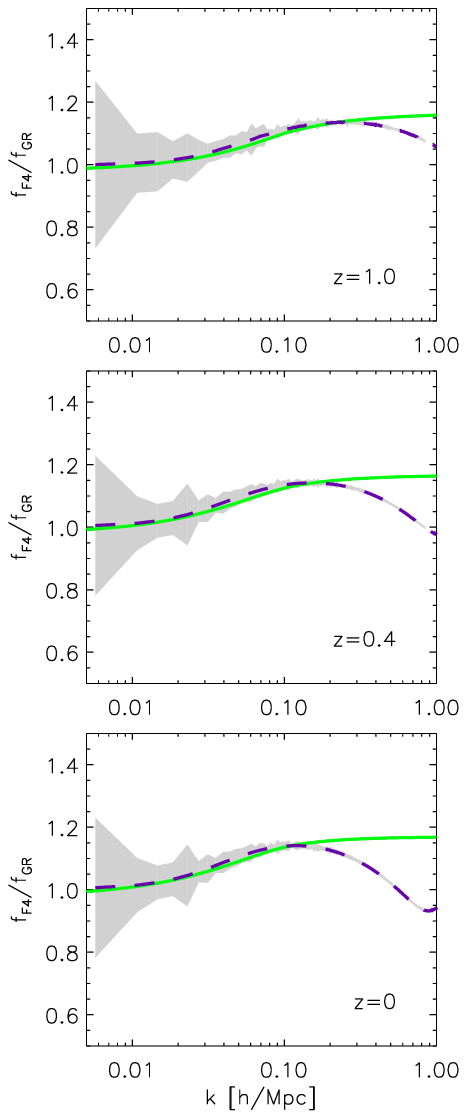}}
\caption{
 The ratio $f_{\rm F4}/f_{\rm GR}$ where 
$f= \sqrt{P_{\theta \theta}/P_{\delta \delta}}$ at $z=0$ (bottom panel), $z=0.4$ (middle panel) and $z=1.0$ (top panel), in the  F4 model compared to GR (purple dashed lines).
The linear theory prediction for this ratio at each redshift is shown as a solid green line.
\label{fratio}}
\end{figure}

In the middle and bottom rows of Fig~\ref{f4theta}  we plot similar ratios  for the F5 and
F6 models respectively. For these two models the difference in the ratios compared to $\Lambda$CDM is less dramatic 
than for the F4 model. This is to be expected as the linear growth rate for these models only 
differs from that in $\Lambda$CDM by at most 6\% for F5 and $\sim1\%$ for F6 on scales $k<0.1 h/$Mpc. 
Furthermore, the simulation results start to deviate from the linear perturbation prediction earlier than it does for the F4 model. This is because 
the suppression of the fifth force itself is a nonlinear effect and the nonlinearity gets weaker as $|f_{R0}|$ increases, making the 
linear perturbation a better approximation for F4.

Fig.~\ref{f4thetaz1} shows the redshift evolution of the ratio $\sqrt{P_{\theta \theta}/P_{\delta \delta}}  = f $ for the 
GR (black dotted line for linear perturbation prediction and blue dashed line measured from $N$-body simulation) and F4 model 
(green solid and purple dashed lines respectively) 
at $z=0.4$ (lower panel) and
$z=1$ (upper panel). The absolute difference in this ratio for these the two models
 is even more pronounced at higher redshift, for the same reason as discussed in the redshift evolution of $P_2/P_0$ (Fig.~\ref{p2p0f4z1}).

The measured ratio agrees with the linear theory predictions for the growth rate on scales $k<0.07 h/$Mpc for the $f(R)$ model and $k<0.04 h/$Mpc
for $\Lambda$CDM at $z=1$, which is again as expected because linear perturbation is a better approximation at earlier times.

Fig.~\ref{fratio} shows the ratio of $f$ in F4 to that in GR at three different redshifts. We also plot the 
linear theory prediction for the ratio $f_{f(R)}/f_{\rm GR}$ as a green solid line in this figure. 
Linear perturbation theory predicts that the ratio becomes larger at lower redshifts,
which is shown by a small increase in
the green line in Fig. 11 at $k \sim 1h/$Mpc. 
On the other hand, the ratio $f_{f(R)}/f_{\rm GR}$ obtained from 
the simulations remains roughly the same at $k<0.2 h/$Mpc-- at all three redshifts and decreases at increasing redshift on smaller scales due to increased
damping in the F4 model compared to GR. The fractional difference is $\sim12\%$ where 
the growth rate in F4 models peaks and the onset of the increase occurs on scales, which is $k\sim0.2, 0.15$ and $0.09h/$Mpc for $z=1,0.4$ and $0$ respectively. 
This is because the 
damping of the velocity power spectrum due to nonlinearity becomes larger at lower redshifts on small scales, 
compensating the enhancement on large scales. The shift of the onset of the peak towards larger scales at later times merely reflects 
the fact that small scales are affected earlier.

\subsection{Extracting the matter and velocity power spectra \label{reconstructing}}

In this section we investigate if a model for the 2D redshift space power spectrum
can be used to extract the density and velocity power spectra, as a function of scale, at $k<0.1 h/$Mpc. If we were able to measure
both of these power spectra accurately
this would provide us with a measure of
the growth rate of structure, as seen in Fig.~\ref{f4theta}, which may be scale dependent as is the case for the $f(R)$ gravity model.
The  motivation for restricting ourselves to these large scales, $k<0.1 h/$Mpc, is that the
impact of bias and nonlinear damping due to velocity dispersions is expected to be small over this range \citep[see e.g.][]{a2008}.

\begin{figure*}
%{\epsfxsize=14.5truecm
%\epsfbox[71 374 542 622]{pk_k_mu.ps}}
 \epsfig{file=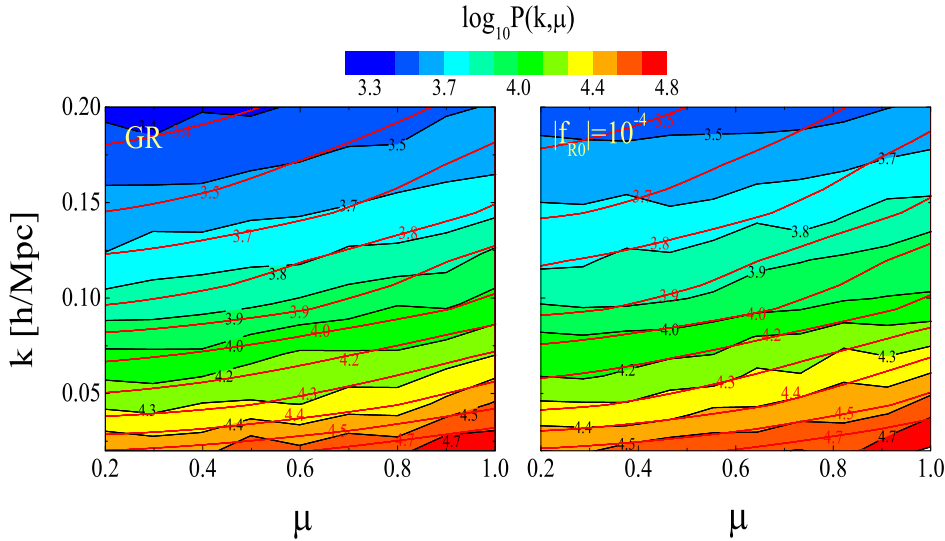,height=3.in,width=6in}
\caption{
The two dimensional redshift space power spectrum, $P(k,\mu)$, measured from the $\Lambda$CDM simulation (left panel) and the F4 simulation (right panel)
 at $z=0$. The coloured contours and black solid lines represent
$\log_{10}P(k,\mu)$. The overplotted red solid lines show the predictions of the model of Eq. \ref{linearmodel}, where the matter and velocity power spectra used 
for each cosmology have been
measured from the simulation.
\label{grpkkmu}}
\end{figure*}

The left and right panels in Fig.~\ref{grpkkmu} show the 2D power spectrum for GR and the F4 model respectively, plotted as a function of wavenumber $k$ and $\mu$ at $z=0$.
The coloured shading represent the values of $\log_{\tiny 10}P(k,\mu)$ measured from the simulations. The overplotted red lines
represent the model of Eq.~\ref{linearmodel} which  uses the velocity and density power spectra measured from each
simulation. From both of these plots it appears that the amplitude predicted by the
model in Eq. \ref{linearmodel} agrees with the measured 2D spectra although it fails to capture the detailed shape of the 2D spectrum over the full range of $k$ and $\mu$. This result agrees with what was found by \citet{kll2012}. 

In order to test the precision with which the model in Eq.~\ref{linearmodel}
can reproduce the 2D power spectrum we shall fit for both the velocity, $P_{\theta \theta}$, and matter power spectra, $P_{\delta \delta}$  under the assumption that in the quasi-linear regime $P_{\delta \theta} = \sqrt{P_{\theta \theta }P_{\delta \delta}}$ %\citep{2009MNRAS.393..297P}.
\citep{pw2009}. We have verified that this is true for our simulations to within a few percent accuracy for $k<0.1 h/$Mpc.
We perform this fit over separate $k$ bins of width $\Delta k = 0.01 h/$Mpc up to a maximum of $k = 0.1 h/$Mpc, using the entire range of $0 < \mu < 1 $.

\begin{figure*}
{\epsfxsize=12.5truecm
\epsfbox[77 363 522 578]{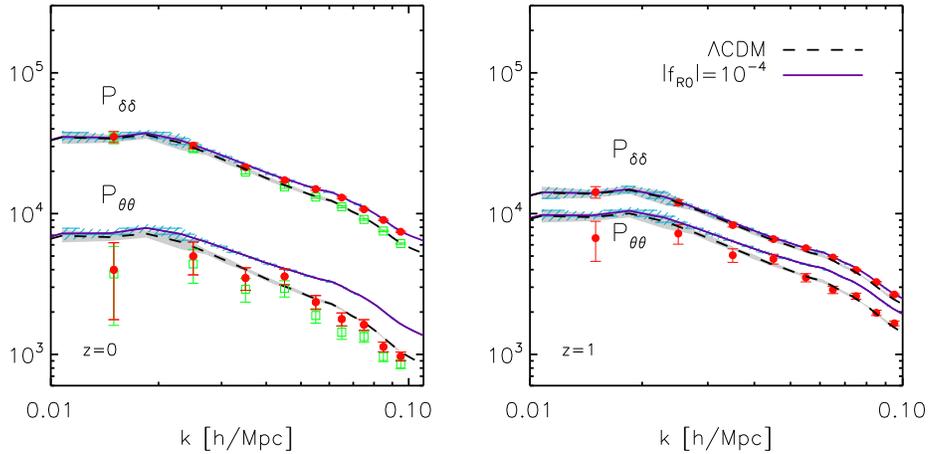}}
\caption{
Left panel: The nonlinear matter (upper curves) and velocity (lower curves) power spectra measured from the
$z=0$ F4 (solid purple) and $\Lambda$CDM (dashed black) simulations. The filled red circles (green squares) show the results from fitting Eq. 25 to the
2D power spectrum, $P(k,\mu)$ from the F4 (GR) simulation.
 The error bars represent the 1-$\sigma$ errors on the fit, solid grey and hatched blue shaded regions represent the errors on the
power spectrum measured from the $\Lambda$CDM and $f(R)$ simulation respectively.
Right panel: Similar to the left panel but for $z=1$.
\label{fitptt}}
\end{figure*}

The results of fitting to the F4 simulation at $z=0$ and $z=1$ are shown in the left and right panels 
respectively in Fig.~\ref{fitptt}. The average power spectra, $P_{\theta \theta}$ (lower curves)
and $P_{\delta \delta}$ (upper curves), measured from the six simulations
are plotted as a black dashed line for $\Lambda$CDM and a solid purple line for the $f(R)$ model at each redshift.
The red filled circles show the results of the fit for each power spectra measured from the F4 simulation.  
At $z=0$ we show the result of the fit for each power spectra measured from the GR simulation as green squares.
At both redshifts it is clear that the
model in Eq. \ref{linearmodel} is able to accurately  describe the amplitude of the 2D $P(k,\mu)$. 
The matter power spectrum, $P_{\delta \delta}$ is recovered accurately and is distinguishable from $\Lambda$CDM.
Unfortunately this model is not able to reproduce the velocity power spectra from the modified gravity model and at both redshifts is
biased to lower values. We have verified that fitting Eq. \ref{linearmodel}
over a reduced range in $\mu$ allows us to  recover the correct $P_{\theta \theta}$ but at the cost of an increase in the errors by more than the difference in the two cosmologies. 
These results demonstrate that the redshift space distortion model $P(k,\mu) = P_{\delta \delta}(k) +  2\mu^2\sqrt{P_{\theta \theta }P_{\delta \delta}} + \mu^4P_{\theta \theta}(k)$, 
accurately describes the amplitude of the 2D power spectrum and can recover $P_{\delta \delta}$ but the angular dependence on $\mu$ is incorrect and so we cannot extract $P_{\theta \theta }$.
These results are in agreement with work by \citet{tang2011}. These authors fit for the density velocity cross power spectrum $P_{\delta \theta }$ 
and find a similar bias in recovering the velocity power spectrum using this model. 
We have also fit the model in Eq. 23 allowing the velocity dispersion damping term to be a free parameter. This method recovers the correct $P_{\theta \theta}$ within the error bars but
the constraints on the velocity power spectra are too weak to distinguish between GR and the F4 model.
This measurement of $P_{\theta \theta}$ would not be accurate enough to allow us to
discriminate between the F4 model and $\Lambda$CDM at either redshift.

\subsection{Modelling the moments of the redshift space $P(k)$ \label {moments}}

In this section we return to studying the moments of the redshift space power spectrum, $P_0$ and $P_2$. 
As shown in Section \ref{reconstructing} the model given in Eq. 25 fails to capture the shape of the full 2D 
$P(k,\mu)$ so naively we do not expect that we can precisely measure these moments on all scales. In this section 
we investigate how well this model works at recovering the measured moments on large scales after averaging over $\mu$.  
Previous work has shown that
the model given in
Eq. \ref{linearmodelmoments} provides a good fit to measurements from simulations on quasi-linear scales
$k<0.3 h/$Mpc and at high redshifts $z\approx 1$, without the need to include a damping term \citep{jbp2011a}. %\citep{2011MNRAS.410.2081J}.

\begin{figure*}
{\epsfxsize=9.5truecm
\epsfbox[76 367 362 651]{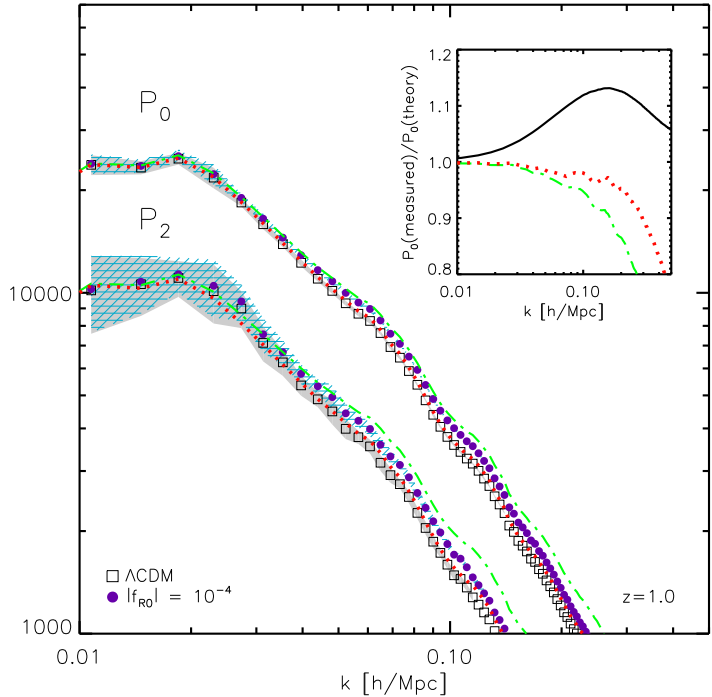}}
\caption{
Upper curves: The monopole moment of the redshift space power spectrum measured from the F4 (filled purple circles) and $\Lambda$CDM (empty
black squares) simulations at $z=1$.
Lower curves: The quadrupole moment of the redshift space power spectrum measured from the F4 (filled purple circles) and $\Lambda$CDM (empty
black squares) simulations at $z=1$.
The two moments have been offset for clarity in this plot.
The model in Eq.~\ref{linearmodel} for each power spectra moment is shown as a dot-dashed green line for the $f(R)$ model and as a red dotted line for $\Lambda$CDM.
The hatched (solid) shaded regions represent the errors on the measured power spectra for the $f(R)$ ($\Lambda$CDM) simulations.
The inset panel shows the ratio of the monopole moment to the model in each cosmology, F4 (dot dashed green lines) and $\Lambda$CDM (red dotted lines).
The solid black line  shows the ratio of the monopole moment in the F4 model to $\Lambda$CDM.
\label{momentsplot}}
\end{figure*}

The $z=1$ multipole moments, $P_0$ (upper curves) and $P_2$ (lower curves), measured from the  $\Lambda$CDM (empty black squares) and
F4 (filled purple circles) simulations are shown
in Fig.~\ref{momentsplot}. These two power spectra have been separated in this plot for clarity. The model given in
Eq.~\ref{linearmodelmoments} using the velocity and matter power spectra
from the simulations are overplotted as a green dot-dashed line and a red dotted line for the $f(R)$ and the $\Lambda$CDM
cosmologies respectively. The inset panel shows the ratio of the measured $P_0$ to the model for each cosmology,  
F4 (dot dashed green lines) and $\Lambda$CDM (red dotted lines).
The model for the monopole moment reproduces the measurement for the $f(R)$ model to within 10\% accuracy at $k<0.2 h/$Mpc.
The $P_0$  model for $\Lambda$CDM is accurate to within 5\% at $k<0.2 h/$Mpc.
The solid black line in the inset panel shows the ratio of the monopole moment in the F4 model to $\Lambda$CDM.
 The model
precision  for $\Lambda$CDM is sufficient to detect the 15\% difference in the monopole moment which we find between the
two cosmologies on these large scales.
The model in Eq. \ref{linearmodelmoments} requires accurate knowledge of the velocity and matter power spectra as input parameters.
In this Section we have used  $P_{\theta \theta}$ and $P_{\delta \delta}$ measured from the simulations.
An alternative to this would be to use fitting formula for each of these power spectra which have
sufficient accuracy on these large scales \citep[see e.g.][]{setal2003, jbp2011a}. %{Smith:2002dz, 2011MNRAS.410.2081J}.

\section{Summary and Conclusions \label{CONCLUSIONS}}

Modified gravity theories generally predict different clustering 
properties of matter, and as a result both the 
density and the velocity power spectra could be very different from the 
predictions of general relativity. The $f(R)$ gravity model has been a leading example of this in recent years. 
Here, the enhancement to the standard gravity depends sensitively on the local matter density through the so-called chameleon mechanism. 
In high matter density and high curvature regions ($f_R\ll|\Phi|$ where $\Phi$ is the Newtonian potential), 
the enhancement is strongly suppressed and standard gravity is recovered; on the other hand, in low matter density and low curvature regions 
the enhancement factor can be as large as $4/3$. Depending on the value of $f_{R0}$ and the local environment, 
the transition scale, or Compton length, of the scalaron ranges from less than one to more than 
a few mega-parsecs, which can potentially leave detectable features in the distribution of matter and clustering patterns of the large-scale structure.

Galaxy surveys measure the distribution of matter in redshift space where the true position of a galaxy appears
 distorted along the line of sight 
due to peculiar velocities.
The goal of many current and furture galaxy redshift surveys is to constrain deviations from GR and so it 
is important to understand how observables are affected by redshift space distortions. 
Theoretical studies of this require both high resolution and large volume 
numerical simulations, which previously have not been performed for 
modified gravity models. 
In this paper we use simulations in large volume boxes to carry out the first study of the clustering of the dark matter in redshift 
space in a $f(R)$ modified gravity. The simulation code developed by \citet{lztk2012} allows 
us to model large volumes with good resolution.

The simulations we use in this analysis 
have two different resolutions, with 1024$^3$ dark matter 
particles in computational boxes of $L_{\rm box}=1.5h^{-1}$Gpc and $1.0h^{-1}$Gpc on a side. 
We have checked that the simulations agree with one another down to $k
\sim0.3h/$Mpc. 
We have compared the matter $P(k)$ measured in real space against previous simulations and found good agreement.

We have measured the redshift space power spectrum in the GR and the $f(R)$ cosmologies
at redshifts $z=0, 0.4$ and 1. We find an enhanced boost in the power on large scales and a substantial increase
in the damping on small scales in the $f(R)$ cosmology compared to GR at all redshifts. The deviations are largest for
the $f(R)$ model with parameter value $|f_{R0}|=10^{-4}$ (F4) and are reduced as $|f_{R0}|$ is decreased.
The large scale enhancement of the power is a result of the fifth force in the modified gravity theory, 
which strengthens the matter clustering on large scales. 
On small scales where the local curvature is not too high, the fifth force makes the particles move faster, 
increasing the velocity dispersion and causing a stronger damping of the 
power compared to GR. 
These effects can be seen at various redshifts, and are even slightly stronger at 
earlier times. 
However, for some model parameters such as $|f_{R0}|=10^{-6}$, the fifth force is strongly suppressed by the 
chameleon mechanism, and its effect is too weak to be distinguished from standard gravity.
We find that the relative difference in the  moments of the power spectrum, $P_2/P_0$, between the F4 $f(R)$ model and GR ranges from
20\% at $z=1$ to 40\% at $z=0$ at $k=0.2h/$Mpc due to the enhanced nonlinear damping on small scales.

 We measure the velocity divergence power spectrum using the DTFE method in both $f(R)$ and 
GR cosmologies and find a large difference between $P_{\theta \theta}$ in the $f(R)$ model compared to GR. 
This difference is much larger than the difference between the non linear matter $P(k)$ in the two models suggesting that
the velocity power spectra is a far more sensitive probe of modified gravity.
We find a large deviation in the ratios
$\sqrt{P_{\theta \theta}/P_{\delta \delta}}$ and $P_{\delta \theta}/P_{\delta \delta}$ between the two models at $ 0.03<k(h/$Mpc)$<0.5$ at $z=0$.
In linear theory these ratios equal the growth rate of structure, $f$, on large scales when the velocity divergence is normalized as
$\theta = \vec{\nabla} \cdot \vec{v}/aH$. Our results show that the measured ratios agree with the linear 
growth rate
 for each cosmology, which is scale {\it dependent} in the case 
of modified gravity, for $k<0.06h/$Mpc at $z=0$.
We find that the relative deviation of the measured nonlinear 
ratio, $\sqrt{P_{\theta \theta}/P_{\delta \delta}}$,
 from the linear prediction between the $f(R)$ model and GR decreases with increasing
 redshift.

Using a simple quasi linear model for the 2D redshift space power spectrum, Eq.~(\ref{linearmodel}), 
which includes nonlinear velocity terms but no small scale damping parameter,
we attempt to extract  the matter and velocity power spectra.
On scales $k<0.1 h/$Mpc we can recover the nonlinear matter power 
spectrum to within a few percent for both the $f(R)$ and the GR cosmology. The model fails to
describe the shape of the 2D power spectrum and we are unable to reconstruct the velocity $P(k)$ accurately.
 The fact that this model recovers the non linear matter power spectrum so precisely
 indicates that this method can be used 
to contrain modified gravity models. 
Our simulation results show that improved theoretical models are required in order to measure the 
velocity power spectrum where there is a large difference in the predicted signal between these two cosmologies.

We show that the same model works very well at fitting the 
first two multipole moments of the redshift space power spectrum 
on large scales, especially
at high redshifts. We are able to match the monopole 
moment to within 3\% for GR and 10\% for the $f(R)$ cosmology on
scales $k<0.2 h/$Mpc at $z=1$.
This difference is smaller than the $15\%$ difference in the F4 and GR models on the same scales.

In this study we have addressed two separate questions. The first considers how well a simple 
model for redshift space distortions, Eq. 25, works at recovering the 
nonlinear matter and velocity $P(k)$ and how well it describes the multipole moments of the redshift space power spectrum.
The matter $P(k)$ and the monopole moment, $P_0$ are recovered accurately and as a result we do not think that 
the failure of the model to decribe the full 2D shape of $P(k,\mu)$ is a serious pitfall for future redshift space analyses. 
Our results point to necessary adjustments needed in the model, such as including non-linear damping terms
for example.
Whether the velocity power spectrum can be extracted using more complicated modelling of the 2D power spectrum, 
and how well it can be measured, is left for future analysis.
The second, two-part, question is which observable, $P(k,\mu)$, $P_0$, $P_2$ etc. shows the largest 
difference between an $f(R)$ and a GR cosmology and is the simple 
redshift space distortion model accurate enough to allow us to measure these differences. We observe the largest 
difference between these two cosmologies in the measured non linear velocity power spectrum on scales $k>0.03h/Mpc$ 
and in the ratio of the multipole moments $P_2/P_0$ on scales $k > 0.2 h/Mpc$. Our results show that an improved model for redshift space distortions 
 is needed in order to extract the velocity power spectrum to a sufficient accuracy to distinguish a $f(R)$ from a GR cosmology.
The differences between the ratios of the multipole moments, $P_2/P_0$, in GR and $f(R)$ are mainly on small scales. We shall address whether or not this difference is present in 
the redshift space clustering of halos and the effects of bias in future work.

To conclude, we find that redshift space distortions in modified 
gravity models have an impact on the clustering of dark matter on large and small scales 
to a level which may be distinguished from general relativity.
 Current redshift space distortion models, which are valid on quasi linear scales, are accurate enough to extract the 
non linear matter $P(k)$ in real space from the measured 2D redshift space power spectrum on large scales, allowing us to constrain $f(R)$ modified gravity.
The large difference between the predicted velocity $P(k)$ in the $f(R)$ and GR cosmology make this a very promising observable with which to test GR provided
that this can be accurately extracted from the redshift space power spectrum.
For certain $f(R)$ parameter values, e.g. $|f_{R0}| = 10^{-6}$, the impact of 
modified gravity on both the matter and velocity fields is not significant and any deviations from GR are restricted  to small scales.
In a follow up paper we
will examine the redshift space distortions in the clustering of halos in these modified gravity models on nonlinear scales.
We will test several nonlinear models for
redshift space distortions to predict how well this observable can constrain $f(R)$ modified gravity in future surveys.

%\tc{
%\begin{itemize}
%\item For the first time, we measured the RSD and the velocity power spectrum from large and high-resolution $f(R)$ simulations;
%\item We find more RSD, more power on large scales and more damping on small scales in $f(R)$ than in GR;
%\item Fitting formula works up to $k<0.1 h$/Mpc;
%\item We find a clear scale-dependent $f$ in $f(R)$;
%\item Observational outlook: For a given redshift survey like BOSS, is it possible to test GR by measuring $f\sigma_8$ in different $k$-bins? Something like that.
%\end{itemize}
%}

\section*{Acknowledgments}

EJ acknowledges the support of a grant from the Simons Foundation, award number 184549. This work was supported in part by the Kavli Institute for Cosmological Physics at the University of Chicago through grants NSF PHY-0114422 and NSF PHY-0551142 and an endowment from the Kavli Foundation and its founder Fred Kavli.
BL is supported by the Royal Astronomical Society and Durham University. GBZ and KK
are supported by STFC grant ST/H002774/1. KK is thankful for support from the ERC and the Leverhulme trust.
The calculations for this paper were performed on the ICC Cosmology Machine, which is part of the DiRAC Facility jointly funded by STFC, the Large Facilities Capital Fund of BIS, and Durham University. We thank Lydia Heck for technical support.

%\bibliographystyle{mn2e}
%\bibliography{mybibliography}

\begin{thebibliography}{}
\bibitem[\protect\citeauthoryear{Angulo {\it et al.}}{2008}]{a2008} Angulo, R.~E. and Baugh, C.~M. and Lacey, C.~G., 2008, MNRAS, 387, 921
\bibitem[\protect\citeauthoryear{Baugh \& Efstathiou}{1994}]{be1994} Baugh C.~M., Efstathiou G., 1994, MNRAS, 270, 183
\bibitem[\protect\citeauthoryear{Bertschinger \& Zukin}{2008}]{bz2008} {Bertschinger}, E. and {Zukin}, P., 2008, PRD, 78, 2
\bibitem[\protect\citeauthoryear{Beutler {\it et al.}}{2011}]{beutler2011} {Beutler}, F., {Blake}, C., {Colless}, M., {Jones}, D.~H., {Staveley-Smith}, L., Poole, G.B., {Campbell}, L., {Parker}, Q., {Saunders}, W., {Watson}, F., 2012, arXiv:1204.4725
\bibitem[\protect\citeauthoryear{Beutler {\it et al.}}{2012}]{beutler2012} {Beutler}, F., {Blake}, C., {Colless}, M., {Jones}, D.~H., {Staveley-Smith}, L., {Campbell}, L., {Parker}, Q., {Saunders}, W., {Watson}, F. , 2011, MNRAS, 416, 3017
\bibitem[\protect\citeauthoryear{Blake~{\it et al.}}{2010}]{betal2010} Blake C.~et~al., 2010, MNRAS, 776
\bibitem[\protect\citeauthoryear{Blake~{\it et al.}}{2011}]{betal2011} Blake C.~et~al., 2011, MNRAS, 415, 2876
\bibitem[\protect\citeauthoryear{Blake~{\it et al.}}{2012}]{betal2012} C.~Blake, S.~Brough, M.~Colless, C.~Contreras, W.~Couch, S.~Croom, D.~Croton and T.~Davis {\it et al.}, 2012, arXiv:1204.3674 
\bibitem[\protect\citeauthoryear{Brax~{\it et al.}}{2008}]{bbds2008} Brax P., van de Bruck C., Davis A.~C., Shaw D.~J., 2008, PRD, 78, 104021
\bibitem[\protect\citeauthoryear{Brax~{\it et al.}}{2011}]{bbdls2011} Brax P., van de Bruck C., Davis A.~C., Li B., Shaw D.~J., 2011, PRD, 83, 104026
\bibitem[\protect\citeauthoryear{Brax~{\it et al.}}{2012}]{bdlw2012} Brax P., Davis A.~C., Li B., Winther H.~A., 2012, PRD submitted; arXiv:1203.4812 
\bibitem[\protect\citeauthoryear{Cantun \& van~de~Weygaert}{2011}]{cv2011} Cautun M.~C., van~de~Weygaert R., 2011, arXiv:1105.0370 
\bibitem[\protect\citeauthoryear{Cole, Fisher \& Weinberg}{1994}]{cfw1994} Cole S., Fisher K.~B., Weinberg D.~H., 1994, MNRAS, 267, 785
\bibitem[\protect\citeauthoryear{Carroll {\it et al.}}{2003}]{carroll2003}  Carroll S. M.,  Duvvuri V.,  Trodden M. and  Turner M.S., 2003, PRD , 70, 043528
\bibitem[\protect\citeauthoryear{Davis { \it et~al.}}{2012}]{dlmw2012} Davis A.~C., Li B., Mota D.~F., Winther H.~A., 2012, ApJ in press; arXiv:1108.3081 
\bibitem[\protect\citeauthoryear{De~Felice \& Tsujikawa}{2010}]{dt2010} De~Felice A., Tsujikawa S., 2010, Living~Rev.~Rel.~13, 3
\bibitem[\protect\citeauthoryear{Ferraro {\it et al.}}{2010}]{fsh2010} {Ferraro}, S. and {Schmidt}, F. and {Hu}, W., 2010, PRD, 83, 6
\bibitem[\protect\citeauthoryear{Gil-Mar{\'{\i}}n {\it et al.}}{2011}]{gil2011} {Gil-Mar{\'{\i}}n}, H. and {Schmidt}, F. and {Hu}, W. and {Jimenez}, R. and {Verde}, L., 2011, JCAP, 11, 19
\bibitem[\protect\citeauthoryear{Green {\it et~al.}}{2011}]{Green} Green J., Schechter P., Baltay C., Bean R., Bennett D., Brown R., et al., 2011, Wide-Field
InfraRed Survey Telescope (WFIRST) Interim Report. ArXiv e-prints
\bibitem[\protect\citeauthoryear{Guzzo {\it et al.}}{2008}]{Guzzo} Guzzo, L., et al., 2008, Nature, 451, 541
\bibitem[\protect\citeauthoryear{Hamilton}{1998}]{hamilton1998} Hamilton A.~J.~S., 1998, ASSL, 231, 185
\bibitem[\protect\citeauthoryear{Hu \& Sawicki}{2007}]{hs2007} Hu W., Sawicki I., 2007, PRD, 76, 064004
\bibitem[\protect\citeauthoryear{Jackson}{1972}]{jackson1972} Jackson J., 1972, MNRAS, 156, 1P
\bibitem[\protect\citeauthoryear{Jennings, Baugh \& Pascoli}{2011a}]{jbp2011a} Jennings E., Baugh C.~M., Pascoli S., 2011a, MNRAS, 410, 2081
\bibitem[\protect\citeauthoryear{Jennings, Baugh \& Pascoli}{2011b}]{jbp2011b} Jennings E., Baugh C.~M., Pascoli S., 2011b, ApJ, 727, L9
\bibitem[\protect\citeauthoryear{Kaiser}{1987}]{kaiser1987} Kaiser N., 1987, MNRAS, 227, 1
\bibitem[\protect\citeauthoryear{Khoury \& Weltman}{2004}]{kw2004} Khoury J., Weltman A, 2004, PRD, 69, 044206
\bibitem[\protect\citeauthoryear{Kwan, Lewis \& Linder}{2012}]{kll2012} Kwan J., Lewis G.~F., Linder E.~V., 2012, ApJ, 748, 78
\bibitem[\protect\citeauthoryear{Lombriser {\it et al.}}{2010}]{lssh2010} {Lombriser}, L. and {Slosar}, A. and {Seljak}, U. and {Hu}, W., 2010, arXiv:1003.3009 
\bibitem[\protect\citeauthoryear{Laureijs {\it et al.}}{2011}]{Laureijs} Laureijs R., Amiaux J., Arduini S., Augu`res J. , Brinchmann J., Cole R., et al., Euclid Definition Study Report. ArXiv e-prints
\bibitem[\protect\citeauthoryear{Li \& Barrow}{2007}]{lb2007} Li B., Barrow J.~D., 2007, PRD, 75, 084010
\bibitem[\protect\citeauthoryear{Li \& Barrow}{2011}]{lb2011} Li B., Barrow J.~D., 2011, PRD, 83, 024007
\bibitem[\protect\citeauthoryear{Li \& Hu}{2011}]{lh2011} {Li}, Y. and {Hu}, W., 2011, PRD, 84, 084033
\bibitem[\protect\citeauthoryear{Li \& Zhao}{2009}]{lz2009} Li B., Zhao H., 2009, PRD, 80, 044027
\bibitem[\protect\citeauthoryear{Li \& Zhao}{2010}]{lz2010} Li B., Zhao H., 2010, PRD, 81, 104047
\bibitem[\protect\citeauthoryear{Li~{\it et al.}}{2012}]{lztk2012} Li B., Zhao G., Teyssier R., Koyama K., 2012, JCAP, 1201, 051
\bibitem[\protect\citeauthoryear{LSST Science Collaborations}{2009}]{LSST} LSST Science Collaborations, Dec. 2009. LSST Science Book, Version 2.0. ArXiv e-prints.
\bibitem[\protect\citeauthoryear{Matsubara}{2008}]{matsubara2008} Matsubara~T., 2008, PRD, 78, 083519
\bibitem[\protect\citeauthoryear{Marulli {\it et al.}}{2012}]{mbm2012}{Marulli}, F. and {Baldi}, M. and {Moscardini}, L., 2012, MNRAS, 420, 2377
\bibitem[\protect\citeauthoryear{Mortonson {\it et al.}}{2009}]{mhh2009} {Mortonson}, M.~J. and {Hu}, W. and {Huterer}, D., 2009, PRD, 79, 023004
\bibitem[\protect\citeauthoryear{Mota \& Shaw}{2007}]{ms2007} Mota D.~F., Shaw D.~J., 2007, PRD, 75, 063501
\bibitem[\protect\citeauthoryear{MPgrafic}{}]{mpgrafic} http://www2.iap.fr/users/pichon/mpgrafic.html
\bibitem[\protect\citeauthoryear{Navarro \& Van Acoleyen}{2007}]{nv2007} Navarro I., Van Acoleyen K., 2007, JCAP, 02, 022
\bibitem[\protect\citeauthoryear{Nojri \& Odintsov}{2003}]{no2003}  Nojiri S. and  Odintsov S. D., PRD, 68, 123512
\bibitem[\protect\citeauthoryear{Okumura \& Jing }{2011}]{o2011}  Okumura, T. \& Jing, Y. P., 2011, ApJ, 726, 5
\bibitem[\protect\citeauthoryear{Oyaizu}{2008}]{oyaizu2008} Oyaizu H., 2008, PRD, 78, 123523
\bibitem[\protect\citeauthoryear{Oyaizu, Lima \& Hu}{2008}]{olh2008} Oyaizu H., Lima M., Hu W., 2008, PRD, 78, 123524
\bibitem[\protect\citeauthoryear{Peacock \& Dodds}{1994}]{pd1994} Peacock J.~A., Dodds S.~J., 1994, MNRAS, 267, 1020
\bibitem[\protect\citeauthoryear{Peacock {\it et al.} }{2001}]{p2001} Peacock J.~A. {\it et al.}, 2001, Nature, 410, 169
\bibitem[\protect\citeauthoryear{Peebles }{1976}]{p1976} Peebles, P. J. E., 1976, Ap\&SS, 45, 3
\bibitem[\protect\citeauthoryear{Percival {\it et al.}}{2007}]{Percival:2007yw} Percival W.~J., et al., 2007, MNRAS, 381, 1053
\bibitem[\protect\citeauthoryear{Percival \& White}{2009}]{pw2009} Percival W.~J., White M., 2009, MNRAS, 393, 297
\bibitem[\protect\citeauthoryear{Pueblas \& Scoccimarro}{2009}]{ps2009} Pueblas S., Scoccimarro R., 2009, PRD, 80, 043504
\bibitem[\protect\citeauthoryear{Raccanelli}{2010}]{Raccanelli} Raccanelli, A., Samushia, L., Percival, W.~J., MNRAS, 409, 1525
\bibitem[\protect\citeauthoryear{Reid \& White}{2011}]{rw2011} Reid B.~A., White M., 2011, MNRAS, 417, 1913
\bibitem[\protect\citeauthoryear{Reid {\it et al.}}{2012}]{retal2012} Reid B.~A., et al., arXiv:1203.6641 .
\bibitem[\protect\citeauthoryear{Samushia {\it et al.}}{2012}]{Samushia} Samushia, L., Percival, W.~J.,  Raccanelli, A., 2012, MNRAS, 420, 2102
\bibitem[\protect\citeauthoryear{S{\'a}nchez and DES Collaboration}{2010}]{DES} {S{\'a}nchez}, E. and {DES Collaboration}, 2010, Journal of Physics Conference Series, 259, 012080
\bibitem[\protect\citeauthoryear{S{\'a}nchez {\it et al.} }{2009}]{s2009} S{\'a}nchez, A.~G., {Crocce}, M., {Cabr{\'e}}, A., {Baugh}, C.~M. and {Gazta{\~n}aga}, E., 2009, MNRAS, 400, 1643
\bibitem[\protect\citeauthoryear{Schaap \& van~de~Weygaert}{2000}]{sv2000} Schaap W.~E., van~de~Weygaert R., 2000, A~\&~A, 363, L29
\bibitem[\protect\citeauthoryear{Schlegel {\it et al.}}{2007}]{BOSS} Schlegel, D.~J., et al., 2007, AAS, 38, 132
\bibitem[\protect\citeauthoryear{Schlegel {\it et al.}}{2009}]{Schlegel2009} Schlegel, D.~J., et al., 2009, arXiv:0904.0468 
\bibitem[\protect\citeauthoryear{Schmidt }{2009}]{schmidt2009} Schmidt F., 2009, PRD, 80, 043001
\bibitem[\protect\citeauthoryear{Schmidt { \it et~al.}}{2009a}]{sloh2009} Schmidt F., Lima M., Oyaizu H., Hu W., 2009, PRD, 79, 083518
\bibitem[\protect\citeauthoryear{Schmidt{ \it et~al.}}{2009b}]{svh2009} Schmidt F., {Vikhlinin}, A., Hu W., 2009, PRD, 8, 083505
\bibitem[\protect\citeauthoryear{Scoccimarro}{2004}]{scoccimarro2004} Scoccimarro R., 2004, PRD, 70, 083007
\bibitem[\protect\citeauthoryear{Scoccimarro, Couchman \& Frieman}{1999}]{scf1999} Scoccimarro R., Couchman H.~M.~P., Frieman J.~A., 1999, ApJ, 517, 531
\bibitem[\protect\citeauthoryear{Seljak \& McDonald}{2011}]{seljak2011} {Seljak}, U. and {McDonald}, P., 2011, JCAP, 11, 39
\bibitem[\protect\citeauthoryear{Sotiriou \& Faraoni}{2010}]{sf2010} Sotiriou T.~P., Faraoni V., 2010, Rev.~Mod.~Phys., 82, 451
\bibitem[\protect\citeauthoryear{Smith~{\it et al.}}{2003}]{setal2003} Smith R.~E.~et~al., 2003, MNRAS, 341, 1311
\bibitem[\protect\citeauthoryear{Springel}{2005}]{springel} Springel, V., 2005, MNRAS, 364, 1105
\bibitem[\protect\citeauthoryear{Taruya~{\it et al.}}{2010}]{t2010} Taruya, A., Nishimichi, T \& Saito, S, 2010, Phys. Rev. D, 82, 063522
\bibitem[\protect\citeauthoryear{Tang~{\it et al.}}{2011}]{tang2011} Tang, J., Kayo, I. \& Takada, M.,  2011, MNRAS, 416, 2291
\bibitem[\protect\citeauthoryear{Teyssier}{2002}]{ramses} Teyssier R., 2002, Astron.~Astrophys., 385, 337
\bibitem[\protect\citeauthoryear{Vanderveld ~{\it et al.}}{2012}]{v2012} {Vanderveld}, R.~A. and {Mortonson}, M.~J. and {Hu}, W. and {Eifler}, T., 2012, arXiv:1203.3195
\bibitem[\protect\citeauthoryear{van de Weygaert \& Schaap}{2009}]{vis2009} van de Weygaert, R. \& Schaap, W. 2009, Lecture Notes in Physics, Berlin Springer Verlag,  Vol. 665, The Cosmic Web: Geometric Analysis, ed. V. J. Martınez, E. Saar, E. Martınez-Gonzalez, \& M.-J. Pons-Borderıa, 291–413
\bibitem[\protect\citeauthoryear{Weinberg {\it et al.}}{2012}]{Weinberg:2012es} Weinberg D.~H., Mortonson M.~J., Eisenstein D.~J., Hirata C., Riess A.~G. and Rozo E., 2012, arXiv:1201.2434 
\bibitem[\protect\citeauthoryear{Will}{2006}]{will2006} Will C.M., 2006, Living Reviews in Relativity, 9, 3 
\bibitem[\protect\citeauthoryear{Yoshida, Sheth \& Diaferio}{2001}]{ysd2001} Yoshida N., Sheth R.~K., Diaferio A., 2001, MNRAS, 328, 669
\bibitem[\protect\citeauthoryear{Zhao, Li \& Koyama}{2011}]{zlk2011} Zhao G., Li B., Koyama K., 2011, PRD, 83, 044007
\end{thebibliography}

\bsp

\label{lastpage}

\end{document}